\documentclass{jfm}

\usepackage[load=abbr]{siunitx}	% for displaying si units correctly
\DeclareSIUnit{\wtpercent}{wt.\%}
\DeclareSIUnit{\volpercent}{vol.\%}
\DeclareSIUnit\atm{atm}

\begin{document}

\newtheorem{lemma}{Lemma}
\newtheorem{corollary}{Corollary}

\shorttitle{Spreading volatile binary droplets} %for header on odd pages
\shortauthor{A. G. L. Williams et al.} %for header on even pages

\title{Spreading and retraction dynamics of sessile evaporating droplets comprising volatile binary mixtures}

\author
 {
 A. G. L. Williams\aff{1},
  G. Karapetsas\aff{2},
  D. Mamalis\aff{3},
  K. Sefiane\aff{1},
  O. K. Matar\aff{4}
  \and 
  P. Valluri\aff{1}
  \corresp{\email{prashant.valluri@.ed.ac.uk}}
  }

\affiliation
{
\aff{1}
Institute for Multiscale Thermofluids, School of Engineering, University of Edinburgh, Edinburgh EH9 3FB, UK
\aff{2}
Department of Chemical Engineering, Aristotle University of Thessaloniki, Thessaloniki 54124, Greece
\aff{3}
Institute for Materials and Processes, School of Engineering, University of Edinburgh, Edinburgh EH9 3FB, UK
\aff{4}
Department of Chemical Engineering, Imperial College London, Kensington, London SW7 2AZ, UK
}

\maketitle

\begin{abstract}
The dynamics of thin volatile droplets comprising of binary mixtures deposited on a heated substrate are investigated. Using lubrication theory, we develop a novel one-sided model to predict the spreading and retraction of an evaporating sessile axisymmetric droplet formed of a volatile binary mixture on a substrate with high wettability. A thin droplet with a moving contact line is considered, taking into account the variation of liquid properties with concentration as well as the effects of inertia. The parameter space is explored and the resultant effects on wetting and evaporation are evaluated. Increasing solutal Marangoni stress enhances spreading rates in all cases, approaching those of superspreading liquids. To validate our model, experiments are conducted with binary ethanol-water droplets spreading on hydrophilic glass slides heated from below. The spreading rate is quantified, revealing that preferential evaporation of the more volatile component (ethanol) at the contact line drives superspreading, leading in some cases to a contact line instability. Good qualitative agreement is found between our model and experiments, with quantitative agreement being achieved in terms of spreading rate.
\end{abstract}

\section{Introduction}
A sessile droplet evaporating from a solid substrate is central to a wide variety of processes. Examples range from spray cooling of microelectronics \citep{Bar-Cohen2006,Kim2007,Deng2011} to inkjet printing \citep{Calvert2001,Singh2010}, pesticide deposition \citep{Yu2009,Damak2016} and even disease diagnosis \citep{Sefiane2010a,Brutin2011,Chen2016}. An evaporating sessile droplet is rarely at true equilibrium with the limiting mechanism in non-volatile liquids tending to be the diffusion of vapour away from the interface \citep{Bourges-Monnier1995,Hu2002}. More volatile droplets, however, can be modelled using kinetic theory and interface non-equilibrium effects \citep{Anderson1995,Ajaev2005}. 

Depending on wettability, droplets can either spread completely over the substrate, forming a pancake with a zero contact angle, or they can become pinned at the triple contact line (where solid, liquid, and gas meet), settling at an equilibrium contact angle. In both cases, once spreading is finished, evaporation soon takes over and droplet profile changes, making the non-equilibrium nature of the problem clear. Wettability of a droplet over a substrate can be explained by equation \ref{Young eqn}---the well known Young's equation,
\begin{equation}
\sigma_{SV} - \sigma_{SL} - \sigma_{LV}\cos{\theta_{eq}} = 0
\label{Young eqn}
\end{equation}
where $\sigma$ denotes free energy per unit length (or surface tension) and subscripts $S$, $L$, $V$, refer to the solid, liquid, and vapour respectively. For a partial wetting droplet with a non-zero equilibrium contact angle, the cohesive forces of $\sigma_{SL}$ and $\sigma_{LV}$ are larger than the adhesive force of $\sigma_{SV}$, i.e., $\sigma_{SV} < \sigma_{SL} + \sigma_{LV}$. Therefore, the surface energy is minimised by inward motion of the droplet and results a finite contact angle. For a completely wetting droplet with zero contact angle ($\theta_{eq} = 0$), a special case arises from the fact that $\cos\theta_{eq} = 1$, yielding; $\sigma_{SV} = \sigma_{SL} + \sigma_{LV}$. and so the cohesive and adhesive forces are perfectly balanced.  

Further complexity arises due to the larger number of factors governing sessile droplet dynamics. Behaviour is heavily influenced by properties of the solid substrate, including substrate roughness \citep{Cazabat1986,Nakae1998,Chen2005} and conductivity \citep{Ristenpart2007,Dunn2009a}; the liquid, including surface tension and volatility \citep{Sefiane2008,Starov2009a}; and the surrounding gas, including atmospheric pressure \citep{Sefiane2009a}, humidity \citep{Fukatani2016} and vapour properties \citep{Shahidzadeh-Bonn2006}. In addition, the dynamics are strongly dependent on the temperature of each phase \citep{Girard2008,Sobac2012,Parsa2015}, droplet shape \citep{Saenz2015}, and gravity becomes important as volume increases \citep{Extrand2010,Srinivasan2011}. 

Introduction of miscible and/or immiscible liquids \citep{Christy2011,Bennacer2014,Tan2016} complicates matters even further. For droplets close to or below the capillary length ($L_c = \sqrt{\sigma/\rho g}$), the well known Marangoni effect has a strong influence on the flow field, dictating much of their behaviour \citep{Deegan1997a,Deegan2000}. Correctly identified by Italian physicist Carlo Marangoni, such flows arise due to surface tension gradients owing to both variations in temperature and liquid composition \citep{Scriven1960a}---know as thermal and solutal Marangoni flow respectively. 

The solutal Marangoni effect causes droplets comprising of binary mixtures to display distinctly different behaviours from the single component equivalent. Early work by \citet{Sefiane2003} found that pinned binary droplets of ethanol-water mixtures displayed non-monotonous behaviour, heavily influenced by the initial concentration. This was unlike pure droplets which displayed a monotonous evolution of evaporation rate and interface profile in time \citep{Picknett1977}. The internal flow field of ethanol-water droplets has been shown to be inherently more complex and chaotic \citep{Christy2010,Christy2011} due to surface tension differences arising from the uneven concentration as a result of preferential ethanol evaporation. With these early studies confined to axisymmetric droplets, \citet{Saenz2017} investigated well defined non-spherical geometries and found that controlling the interface curvature would cause segregation of the two components. With evaporation proceeding slowest at areas of minimum curvature, ethanol would linger in these areas for the longest times.

An important study on wetting binary droplets by \citet{Guena2007b} found the remarkable behaviour that binary alkane mixtures tended to spread and evaporate faster than either of their pure constituents---as studied by \citet{Cachile2002a,Cachile2002}. \citet{Guena2007b} noted that spreading would deviate from Tanner's law, with the spreading exponent rising to $n = 0.3$ ($r \propto t^n$). This behaviour was owing to the solutal Marangoni effect. Mixtures were carefully selected so that the less volatile component (LVC) of the mixture had a higher surface tension than the more volatile component (MVC). The preferential evaporation of MVC at the contact line would leave a higher concentration of LVC and hence a higher surface tension compared to the bulk. The surface tension gradient would induce Marangoni flows towards the contact line, enhancing the capillary force and, as a result, the spreading rate. Droplets would spread to minimum thickness more quickly than their single components counterparts and reach dry-out faster, even when only LVC remained, due to the thinner droplet profile and increased interfacial surface area enhancing evaporation. Depending on the initial concentration, interesting drying profiles were observed, such as the droplet centre drying out before the contact line, leaving a torus shaped ring. 

The first complete model to simulate the evaporation of a multicomponent droplet was provided by \citet{Diddens2017a} who extended the mathematical model of \citet{Siregar2013}, based on the lubrication approximation and solved using the finite volume method. They considered partially wetting binary droplets of ethanol-water and water-glycerol evaporating from an isothermal substrate at contact angles 6.6$^\circ$-40$^\circ$ using a Navier-slip condition at the contact line. For ethanol-water droplets, \citet{Diddens2017a} observed that at long times ethanol had almost entirely evaporated but a strong thermal Marangoni flow was still present---validating the hypothesis of \citet{Christy2011}. They noted that when the droplet becomes flat, the surface tension gradient leads to shape deformation with a depression in the droplet centre---similar to the observations of \citet{Guena2007b}. Entrapped residual ethanol, previously predicted \citep{Sefiane2008a,Liu2008}, could not be noticed, which the authors argue was due to strong convective mixing resulting from the fast Marangoni flow. However, residual amounts of water in glycerol-water droplets (where diffusive transport is slower) were found to remain in the later stages. By then extending the model to non-isothermal heated substrates, \citet{Diddens2017a} was able to reproduce the flow regimes and transitions reported experimentally by \citet{Zhong2016}. \citet{Diddens2017b} also approached the problem using a finite element model to tackle larger contact angles above 90$^\circ$, no longer invoking the lubrication approximation.  Thermal convection was also included, accounting for the effects of substrate thickness and evaporative cooling. Here the results showed that the evaporation of the MVC can drastically decrease the interface temperature, causing the the ambient vapour of the LVC to condense onto the droplet. The approach used by \citet{Diddens2017b} was compared with the previous lubrication-based model \citep{Diddens2017a}. While the volume evolutions agreed well, even at low contact angles, the lubrication approach over-predicted the regular Marangoni velocities and under-predicted the chaotic velocities in the case of an instability.

The evaporation of a ternary mixture droplet was investigated for the first time by \citet{Tan2016}. Specifically, partially wetting droplets of the alcoholic beverage, Ouzo---a mixture of water, ethanol, and anise oil. The addition of anise oil adds a further complication of mutual solubility, with the oil being miscible in ethanol but immiscible in water. The evaporation phenomena was revealed to be extremely rich, with evaporation-induced phase separation being observed. \citet{Li2018} also recently observed component segregation in binary droplets due to evaporation from the contact line rim being faster than the induced Marangoni flow, resulting in the convection usually caused by Marangoni flows too weak to maintain perfect mixing.  

From the short review above, while some aspects of evaporating binary mixture droplets have been reported, the underlying physics of spreading (and retraction) dynamics is still in question. This is particularly important for many applications including cooling and development of self-cleaning solvent mixtures that rely on the volatilities. In this paper, we present comprehensive lubrication modelling supported by experiments considering ideal ethanol-water mixtures, far away from azeotropic concentrations. We particularly focus on flat droplets formed due to an underlying hydrophilic substrate. This allows us to not only validate our lubrication model but also to identify spreading regimes whilst at the same time revealing the governing physics. Our simulations elucidate the role of thermal and solutal Marangoni stresses and capillary forces at various stages of the evaporating process. In line with our experimental observations reported herein, it is demonstrated that for a sufficiently high concentration of ethanol, solutal Marangoni stresses drive very fast spreading of the droplet at early stages of evaporation, with spreading exponents that may exceed the value of 1. The enhanced spreading may also be accompanied by the formation of a ridge near the contact line. This behaviour is clearly reminiscent of superspreading reported in surfactant-laden flows \citep{Rafai2002, Karapetsas2011}. As it will be shown below, enhanced spreading of binary mixture droplets is due to the presence of strong Marangoni stresses near the contact line, arising due to the preferential evaporation of ethanol in that region. In contrast to the surfactant laden flows however, the concentration gradients here arise as natural consequence of the evaporation process. At later stages, it is shown that the dynamics of the evaporation and droplet shape is dictated by the interplay of thermal and solutal Marangoni stresses and capillary forces. 

\section{Problem statement and model formulation}\label{model definition}
\subsection{Description of the problem}
\label{Description of the problem}
We study the behaviour of a small and thin sessile droplet consisting of a mixture of two volatile, miscible liquids $A$ and $B$. Liquid $A$ is the more volatile component (MVC) in the mixture and liquid $B$ the less volatile component (LVC). The mixture is assumed to be ideal and the droplet is considered Newtonian with density $\hat{\rho}$, specific heat capacity $\hat{c}_p$, thermal conductivity $\hat{k}$, and viscosity $\hat{\mu}$. For simplicity, and because liquids with similar densities will be chosen for components $A$ and $B$, we assume the liquid mixture to be incompressible and the density of both components equal, such that $\hat{\rho}_A=\hat{\rho}_B=\hat{\rho}$. With the exception of density, the remaining properties vary locally with concentration. We account for this using the following rule of mixtures, shown for generic variable $\hat{\zeta}$ as,
\begin{equation}
\hat{\zeta} = \chi_A\hat{\zeta}_A + (1 - \chi_A)\hat{\zeta}_B 
\end{equation}
where $\chi_A$ is the mass fraction of component $A$ in the mixture (hence $\chi_B = 1 - \chi_A$), while $\hat{\zeta}_A$ and $\hat{\zeta}_B$ denote property values of pure component $A$ and $B$ respectively. Within the liquid mixture, we consider only Fick's Law, with the effects of thermodiffusion arising from the Soret effect neglected. At the interface, the surface tension, $\hat{\sigma}$, of the binary mixture has a linear dependence on both the local concentration of each component and the local temperature, $\hat{T}$, taking the form,
\begin{equation}
\label{Hertz-Knudsen}
\hat{\sigma} = \chi_A ( \hat{\sigma}_{A,r} + \hat{\gamma}_{T,A} (\hat{T} - \hat{T}_r) ) 
+ (1 - \chi_A) ( \hat{\sigma}_{B,r} + \hat{\gamma}_{T,B} (\hat{T} - \hat{T}_r) ) 
\end{equation} 
where $\hat{\gamma}_{T,i} = \partial\hat{\sigma}_{T,i}/\partial\hat{T}$ is the temperature coefficient of surface tension of component $i$ ($i = A, B$). $\hat{\sigma}_{i,r}$ is the surface tension of component $i$ at reference temperature $\hat{T}_r$. We assume this to be the temperature of the vapour phase, $\hat{T}_r = \hat{T}_g$.

The droplet resides on heated horizontal solid substrate kept at a constant temperature $\hat{T}_w$ and is released into a thin precursor film consisting solely of the LVC. Evaporation in the film is stabilised by the disjoining pressure which accounts for the attractive van der Waals interactions. The inclusion of the precursor film removes the stress singularity that can arise at the moving contact line. Rather than a purely artificial tool, the precursor film is also a physical effect with experimental verification \citep{DeGennes1985}. The precursor film is always formed on the solid surface if the droplet is surrounded by its vapour, from which it is adsorbed. The precursor film is sufficiently thin that the liquid molecules are attracted to the substrate by van der Waals interactions, stabilising the film and suppressing evaporation \citep{Ajaev2005, Berthier2013}.
\begin{figure}
	\includegraphics[width=\textwidth]{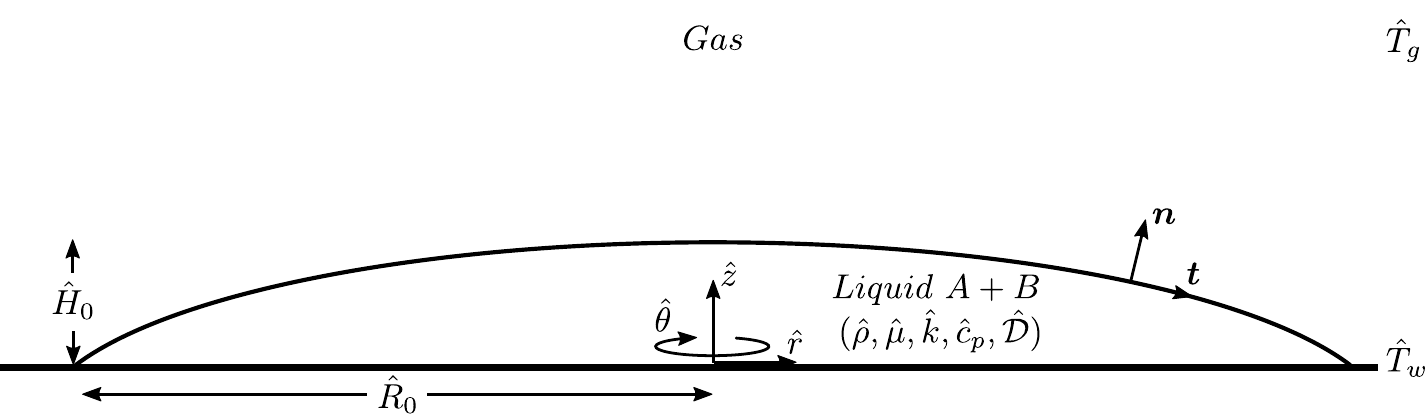}
	\caption{Droplet geometry of initial height $\hat{H}_0$ and radius $\hat{R}_0$ in the cylindrical coordinate frame. The droplet consisting of miscible components $A$ and $B$ and resides on a heated substrate at temperature $\hat{T}_w$. The droplet is sufficiently thin such that the aspect ratio is much less that unity, ${\hat{H}_0}/{\hat{R}_0} \ll 1$. Gas temperature is kept constant at $\hat{T}_g$. $\boldsymbol{n}$ and $\boldsymbol{t}$ denote the outward units vectors acting in normal and tangential directions to the interface respectively.}
\end{figure}

The droplet is in contact with the gas phase which has a bulk temperature of $\hat{T}_g$. The velocity of the gas and vapour particles are assumed sufficiently low so that is negligible. The gas phase has density $\hat{\rho}_v$, viscosity $\hat{\mu}_v$ and thermal conductivity $\hat{k}_v$. These gas-phase properties are assumed to be significantly smaller than their liquid counterparts, such that,  $\hat{\rho}_g \ll \hat{\rho}$, $\hat{\mu}_v \ll \hat{\mu}$, $\hat{k}_v \ll \hat{k}$ \citep{Burelbach1988}. The same is assumed for the vapour properties. In addition, we assume that the total gas phase pressure is sufficiently large that it remains constant with evaporation and changing vapour pressure. 

Given these assumptions, we adopt the so called `one-sided' model and focus solely on the liquid phase in this study. The draw of such an approach is the considerably reduced complexity by discounting the vapour phase while including the physics of the liquid phase. A clear limitation is that we are forced to assume evaporation is not vapour diffusion limited and instead controlled by the transfer of molecules across the liquid-vapour interface. Physically, we are assuming that vapour diffuses rapidly away from the liquid-vapour interface and therefore the model is expected to be valid in the regime where there is a well mixed environment and so the phase-transition process is the rate limiting step. Phase transition is modelled using the non-equilibrium Hertz-Knudsen relation from kinetic theory \citep{Plesset1976,Moosman1980}, written in dimensional for each for each $i$ component as,
\begin{equation}
\label{mass flux i}
\hat{J}_i = \frac{\hat{p}_{v,i} \hat{M}_i}{\hat{R}_g \hat{T}|_h}\bigg(\frac{\hat{R}_g\hat{T}|_h} {2\pi\hat{M}_i}\bigg)^{\frac{1}{2}} \bigg(\alpha_{v,i}\frac{\hat{p}_{v,e,i}} {\hat{p}_{v,i}} - \beta_{v,i}\bigg)
\end{equation}
where $\hat{p}_{v,i}$ is the partial pressure of component $i$, $\hat{p}_{v,e,i}$ is its equilibrium vapour pressure, and $\hat{M}_i$ its molecular weight. $\hat{T}|_h$ denotes the interfacial temperature of the liquid and $\hat{R}_g$ is the universal gas constant. $\alpha_{v,i}$ and $\beta_{v,i}$ are accommodation coefficients for evaporation and condensation respectively, giving the probability that a molecule of component $i$ impinging on the interface will cross over to the other phase \citep{Knudsen1950}. As reviewed in \citet{Murisic2011}, the value of accommodation coefficients used in the literature varies over several orders of magnitude from $O(10^{-6})$ to $O(1)$, with lower values providing a greater barrier to phase change by reducing the probability of a molecule crossing the interface. For simplicity, and in line with other works \citep{Moosman1980,Ajaev2005,Sultan2005}, we assume in this study that the accommodation coefficients are constant and nearly equal to each other, such that $\alpha_{v,i} = \beta_{v,i} = 1 $. Physically this means there is no barrier to phase change and every molecule of vapour or liquid striking the interface transitions to the opposite phase \citep{Persad2016}.

%For simplicity, and in line with other similar models \cite{Moosman1980,Ajaev2017}, we assume evaporation and condensation coefficients are assumed equal to each other and constant 

%the system is always near equilibrium and that each molecule of vapour or liquid striking the interface changes to the opposite phase \cite{Persad2016}, hence we set $\alpha_{v,i} = \beta_{v,i} = 1$.

%Non-equilibrium one-sided (NEOS) models like these have been used extensively to model heated volatile thin films \citep{Burelbach1988} and droplets \citep{Anderson1995,Ajaev2005}.

Another modelling approach not considered here is the `1.5 sided' or `lens' model; generally used when evaporation is firmly in the vapour-diffusion limited regime. When using this method, the liquid phase is fully resolved with the gas phase being solved for diffusion only and boundary conditions applied along the liquid-vapour interface for the liberation of the liquid to vapour. \citet{Murisic2011} have explored when one evaporation model is more appropriate than the other for pure droplets of either water or isopropanol with moving contact line on non-heated surfaces. They concluded that a NEOS model with a small accommodation coefficient, $\alpha_v$, of  $O(10^{-4})$ better reflected the experimental results for pure water droplets while the lens model was more accurate for the isopropanol droplets. 

By using accommodation coefficients close to unity, we expect our model to over predict the evaporation rates compared to experiment, where the vapour diffusion from the interface to a far-field value is typically several orders of magnitude slower than the liberation of liquid molecules to the vapour phase. In practice, this means while our model will qualitatively simulate evaporation, a quantitative comparison with evaporation fluxes against diffusion-limited experiments is impossible. To achieve a quantitative comparison, a modified accommodation coefficient or more complex models such as those of \citet{Sultan2005} or \citet{Saenz2015} should be explored. Despite this, one-sided models similar to the one considered here have proved powerful in the prediction of qualitative behaviour for evaporating droplets in the past, for example the  prediction of hydrothermal waves in evaporating pure component droplets \citep{Karapetsas2012}.

Initially, we assume that the droplet has maximal thickness $\hat{H}_0$ and radius $\hat{R}_0$, in a polar coordinate system $(\hat{r}, \hat{z}, \hat{\theta})$ representing the radial, axial and azimuthal axes. We consider the droplet to be axisymmetric and very thin. Therefore, $\hat{R}_0 \gg \hat{H}_0$, so that the droplet aspect ratio, $\varepsilon = \hat{H}_0/\hat{R}_0 \ll 1$. This assumption permits the use of lubrication theory, which we will employ to derive the evolution equations. Additionally, we assume the droplet is sufficiently small as to neglect gravitational effects. This means a Bond number of much less than one, requiring the radius of the droplet to be below the capillary length of both liquids in the mixture. A working mixture of ethanol and water is considered. Both liquids are sufficiently volatile on a heated substrate, ethanol being the MVC and possessing a lower surface tension than water. The selection of an ethanol-water mixture also avoids any `self-rewetting' properties \citep{Abe2004} present in other alcohol-water mixtures at certain concentrations, for example butanol-water. The pure component properties of each fluid in the mixture are given in table \ref{ethanol-water table}.
\begin{table}
\begin{center}
\def~{\hphantom{0}}
\setlength{\tabcolsep}{15pt}
\begin{tabular}{lS[table-format=1.3e1]S[table-format=1.3e1]}
                                            &{Ethanol}      &{Water}    \\[3pt]
$\hat{\rho}$ (\si{\kg\per\m\cubed})		    &8.00e2         &9.99e2     \\
$\hat{\mu}$	(\si{\Pa\s})				    &1.198e-3	    &6.513e-4   \\
$\hat{k}$ (\si{\W\per\m\per\K})			    &1.83e-1		&6.02e-1 	\\
$\hat{c}_p$	(\si{\kJ\per\kg\per\K})	    	&2.40	        &4.182 		\\
$\hat{L}_v$	(\si{\kJ\per\kg})		    	&1.030e3		&2.454e3 	\\
$\hat{\sigma}_R$ (\si{\N\per\m})	    	&2.28e-2	    &7.29e-2 	\\
$\hat{\gamma}_T$ (\si{\N\per\m\per\K})	    &8.32e-5	    &1.51e-4 	\\
$\hat{M}$ (\si{\kg\per\mole})			    &4.61e-2	    &1.80e-2	\\
$\hat{p}^o$ (\si{\newton\per\metre\squared})&5.80e3		    &7.37e3 	\\
$\hat{\mathcal{D}_A}$ (\si{\m\squared\per\s})&1.23e-9	&  		        \\
\end{tabular}
\caption{Physical properties of ethanol (MVC) and water (LVC) at 20$^\circ$ and 1 atm.}{\label{ethanol-water table}}
\end{center}
\end{table}
\subsection{Governing equations and boundary conditions}
\subsubsection{Scaling}
All of the aforementioned variables have taken dimensional form---a hat ($\:\hat{}\:$) signifying the dimensional symbol. We scale the system using the properties of the more volatile component (MVC), $A$, and the thermocapillary velocity, defined as $\hat{U} = \varepsilon \hat{\gamma}_l \Delta \hat{T} / \hat{\mu}_l$. As such, we now introduce the following scalings: 
\begin{equation}
\left. \begin{array}{l}
\displaystyle
\hat{r} = \hat{R}_0 r,\quad  
\hat{z} = \hat{H}_0 z,\quad 
\hat{t} = \frac{\hat{R}_0}{\hat{U}} t,\quad 
\mathbf{\hat{u}} = (\hat{u}, \hat{w}) = \bigg( \hat{U} u,\frac{\hat{H}_0}{\hat{R}_0}\hat{U} w \bigg); \\[16pt]
\displaystyle
\hat{p} = \hat{p}_{ig} + \frac{\hat{\mu}_A\hat{U} \hat{R}_0}{{\hat{H}_0}^2} p,\quad 
\hat{T} = \hat{T}_0 + T\Delta\hat{T},\quad 
\hat{J}_i = \frac{\hat{k}_A\Delta\hat{T}}{\hat{H}_0\hat{L}_{v,A}} J_i; \\[16pt]
\displaystyle
\hat{\sigma}_i = \hat{\sigma}_{A,0}\sigma_i,\quad 
\hat{\mu} = \hat{\mu}_A \mu,\quad \hat{k} = \hat{k}_A k,\quad 
\hat{c}_p = \hat{c}_{p,A} c_p.
\end{array} \right\}
\end{equation}
Here, $\hat{t}$ is time, $\hat{p}$ is pressure and $\mathbf{\hat{u}}$ is the velocity vector field with components $\hat{u}$ and $\hat{w}$ in the radial and axial directions, respectively. Also, $\hat{L}_v$ is latent heat of vapourisation, $\hat{J}_i$ is the evaporative flux of component $i$ and $\Delta \hat{T}=\hat{T}_w-\hat{T}_g$. The principal dimensionless numbers arising from the scaling are the Marangoni number, $Ma = \hat{\gamma}_A \Delta\hat{T}/\hat{\sigma}_{A,r}$, the Reynolds number, $Re = \hat{\rho}_A \hat{U} \hat{H}_0 / \varepsilon \hat{\mu}_A$, the Prandtl number, $Pr   = \hat{\mu}_A \hat{C}_{p,A} / \hat{k}_A$, the P\'{e}clet number, $Pe = \hat{U} \hat{R}_0 / \hat{\mathcal{D}}_A$, evaporation number, $E = \hat{k}_A \Delta\hat{T} \hat{R}_0 /{\hat{H}_0}^2 \hat{L}_{v,A} \hat{U} \hat{\rho}$, and the Knudsen number,  $K = \hat{k}_A (2\pi{\hat{R}_g}^3 {\hat{T}_g}^5)^\frac{1}{2} / \hat{H}_0 \hat{L}^2_{v,A} \hat{p}_{s,A} {\hat{M}_A}^\frac{3}{2}$. $K$ measures the importance of kinetic effects at the interface and can be thought of as being analogous to inverse of the Biot number, controlling the heat loss across the interface \citep{Karapetsas2012}. In addition, several property ratios unique to the binary mixture also arise from the scaling:
\begin{equation}
\left. \begin{array}{l}
\displaystyle
\sigma_R = \frac{\hat{\sigma}_{B,r}} {\hat{\sigma}_{A,r}}, \quad
\gamma_R = \frac{\hat{\gamma}_{T,B}} {\hat{\gamma}_{T,A}}, \quad 
\alpha = \frac{\hat{p}_{s,B}} {\hat{p}_{s,A}}, \quad
k_R = \frac{\hat{k}_B} {\hat{k}_A}; \\[16pt]
\displaystyle
\mu_R = \frac{\hat{\mu}_B} {\hat{\mu}_A}, \quad
c_{p R} = \frac{\hat{c}_{p,B}} {\hat{c}_{p,A}}, \quad
M_R = \frac{\hat{M}_B} {\hat{M}_A}, \quad
\Lambda = \frac{\hat{L}_{v,B}} {\hat{L}_{v,A}}. \\[16pt]
\end{array} \right\}
\end{equation}
where $\sigma_R$ is the ratio of surface tensions, $\gamma_R$ is the ratio of surface tension temperature coefficients, $\alpha$ is the relative volatility (not to be confused with  $\alpha_v$ in equation \ref{Hertz-Knudsen}), $k_R$ is the ratio of thermal conductivities, $\mu_R$ is the viscosity ratio, $c_{p R}$ is the ratio of specific heats, $M_R$ is the molar weight ratio, and $\Lambda$ is the ratio of latent heats.
\subsubsection{Dimensionless governing equations}
Flow within the droplet is incompressible and governed by the following mass, momentum, energy and concentration equations:
\begin{equation}
\nabla \cdot \mathbf{u} = 0
\label{dimensionless mass}
\end{equation}
\begin{equation}
\varepsilon Re \bigg( \frac{\partial \mathbf{u}}{\partial t}  + \mathbf{u} \cdot \nabla \mathbf{u} \bigg ) + \nabla p  - \nabla^2 \mathbf{u} = 0
\label{dimensionless mom}
\end{equation}
\begin{equation}
\varepsilon Re Pr \bigg( \frac{\partial (c_p T)}{\partial t} + \mathbf{u} \cdot \nabla (c_p T) \bigg) - \nabla k (\nabla T) = 0
\label{dimensionless eng}
\end{equation}
\begin{equation}
Pe \bigg( \frac{\partial \chi_A} {\partial t} + \nabla \cdot \mathbf{u} \chi_A \bigg) -\nabla^2 \chi_A = 0
\label{dimensionless conc}
\end{equation}
The concentration equation \ref{dimensionless conc} is simplified by applying the limit of weak diffusion and assuming  $Pe \approx O(\varepsilon^{-2})$, as derived by \citet{Matar2002}. Therefore, re-defining $Pe = Pe'\varepsilon^{-2}$ and substitution into equation \ref{dimensionless conc} yields the amended conservation equation for $\chi_A$:
\begin{equation}
\frac{\partial \chi_A} {\partial t} + \nabla \cdot \mathbf{u} \chi_A - \frac{\varepsilon^2}{r} \frac{\partial}{\partial r} \bigg( r \frac{\partial \chi_A}{\partial r}\bigg) - \frac{1}{Pe'}\bigg(\frac{\partial^2 \chi_A}{\partial z^2}\bigg) = 0
\label{dimensionless conc 2}
\end{equation}
Note that contrary to the standard approach of lubrication theory, we do not remove the third term on the LHS, despite $\varepsilon^2 \ll 1$. Retaining this weak diffusive force along $r$ ensures that the concentration profile remains numerically stable as the solution proceeds. We also explored the limit of rapid vertical diffusion and and found no qualitative differences with the simulation presented in this manuscript.

Evaporative effects are modelled using a constitutive equation based on the Hertz-Knudsen expression given by equation \ref{Hertz-Knudsen}, written here in dimensionless form as,
\begin{equation}
K J = \chi_A \Big( \delta p + T \vert _h \Big) + (1-\chi_A) \alpha M^{3/2}_R \Big(\delta p + \Lambda T \vert _h \Big)
\label{dimensionless total mass flux}
\end{equation}
where $T \vert _h$ is the temperature of the interface and $\delta = \hat{\mu}_A \hat{U} \hat{R}_0 \hat{T}_g / \hat{\rho}_l {\hat{H}_0}^2 \hat{L}_{v,A} \Delta\hat{T}$ accounts for the effects of changes in liquid pressure on the local phase change temperature at the interface \citep{Ajaev2005}. We partition equation \ref{dimensionless total mass flux} into two separate expressions, yielding the evaporative fluxes of components $A$ and $B$ respectively,
\begin{equation}
J_A = \frac{\chi_A}{K} \big( \delta p + T\vert_h \big)
\label{mass flux A}
\end{equation}
\begin{equation}
J_B = \frac{(1-\chi_A) \alpha M^{3/2}_R}{K} \big(\delta p + \Lambda T\vert_h \big)
\label{mass flux B}
\end{equation}
\subsubsection{Interfacial boundary conditions}\label{interfacial boundary conditions}
Turning our attention to the remaining interfacial boundary conditions at $z = h(r,t)$, the evaporative flux boundary condition at the interface takes the form,
\begin{equation}
EJ = - (u - u_s)\frac{\partial h}{\partial r} + (w - w_s)
\end{equation}
where $u_s$ and $w_s$ are interface velocities of the liquid and $J$ is the total evaporative flux comprising $J_A + J_B$. The associated energy balance is given as,
\begin{equation}
J_A + J_B\Lambda + k \frac{\partial T}{\partial z} = 0
\end{equation}
Let us now consider briefly the gas phase, consisting of inert gas and the vapour of both components $A$ and $B$. Under Dalton's law, the total gas pressure is written as the sum of the partial pressures of each component,
\begin{equation}
\hat{p}_g = \hat{p}_{ig} + \hat{p}_{v,A} + \hat{p}_{v,B}
\end{equation}
Here, $\hat{p}_{ig}$, $\hat{p}_{v,A}$ and $\hat{p}_{v,B}$ indicate the partial pressures of inert gas, component A and component B, respectively. We assume that the surrounding gas phase consists mainly of inert gas rather than vapour, meaning $\hat{p}_{ig} \gg \hat{p}_{v,A}$ and $\hat{p}_{ig} \gg \hat{p}_{v,B}$. This leads to the simplification that the total gas phase pressure is approximately equal to the pressure of the inert gas,
\begin{equation}
\hat{p}_{ig} \approx \hat{p}_g %\approx (\hat{p}_g - \hat{p}_{v,A}) \approx (\hat{p}_g - \hat{p}_{v,B})
 \label{new gas pressure}
\end{equation}
Additionally, since the droplet is considered to be small, we also ignore the effects of vapour recoil from the gas phase \citep{Larson2014} since this will be relatively weak when compared to the dominating surface tension force. Given these assumptions, the normal stress boundary condition at the interface is defined as,
\begin{equation}
\hat{p} - \hat{p}_g + \frac{\varepsilon^2 \sigma}{Ma} 2\kappa + \frac{\mathcal{A}}{h^3} = 0
\label{normal stress BC}
\end{equation}
where $2 \kappa$ is the mean curvature of the interface and $\mathcal{A} = \hat{\mathcal{A}} / 6 \pi \hat{\mu}_A \hat{U} \hat{R}_0 \hat{H}_0$ is the Hamaker constant, made dimensionless in the disjoining pressure term and accounting for intermolecular interactions near the contact line. The interface height, $h$, is handled via the kinematic boundary condition imposed as,
\begin{equation}
\frac{\partial h} {\partial t} + \mathbf{u} \cdot \nabla h + EJ = 0
\end{equation}
We now consider the concentration boundary condition along the interface by applying the limit of weak diffusion introduced in equation \ref{dimensionless conc 2} above. As outlined in \citet{Matar2002}, we derive an expression independent of $z$ by employing an approximate Galerkin expansion for $\chi_A$, seeking solutions of the form,
\begin{equation}
\chi_A(r,z,t) = \chi_{A0}(r,t) + \chi_{A1}(r,t) \bigg(\frac{z^2}{h^2} - \frac{1}{3} \bigg)
\label{Galerkin expansion Xa}
\end{equation}
where $\chi_{A0}$ corresponds to the mean concentration and $\chi_{A1}$ is a non-zero mean quadratic fluctuating component. The concentration balance over the interface is given as,
\begin{equation}
\bigg[ \frac{\partial \chi_{A}}{\partial z} \bigg]_{h} = E(\chi_{A} J - J_A)
\label{weak diffusion BC z=h}
\end{equation}
Differentiation of equation \ref{Galerkin expansion Xa} w.r.t. $z$ and evaluation at the interface ($z = h$) gives an alternative expression for $[{\partial \chi_A}/{\partial z}]_{h}$ in terms of $\chi_{A1}$,
%
%\begin{equation}
%\frac{\partial \chi_A}{\partial z} = \bigg( \frac {\chi_{A1}}{h^2} \bigg) 2 z
%\end{equation}
%
%
\begin{equation}
\bigg[\frac{\partial \chi_A}{\partial z} \bigg]_{h} = \frac{2 \chi_{A1}}{h}
\label{dXa/dz z=h definition}
\end{equation}
Substitution of equation \ref{weak diffusion BC z=h} into \ref{dXa/dz z=h definition} hence constructs an expression for $\chi_{A}$ in terms of $\chi_{A1}$,
\begin{equation}
\chi_A = \frac{2 \chi_{A1}}{E J h} + \frac{J_A}{J}
\label{Xa expression weak diffusion}
\end{equation}
By evaluating equation \ref{Galerkin expansion Xa} at $z = h$ and substituting in \ref{Xa expression weak diffusion}, we obtain the following expression for $\chi_{A1}$ independent of $\chi_A$,
\begin{equation}
\chi_{A1} = \frac{(J_A - J \chi_{A0})}{2 \big( \frac{J}{3} - \frac{1}{Pe' E h} \big)}
\label{Xa1 expression weak diffusion}
\end{equation}
We arrive at the final form of the concentration balance over the interface in the limit of weak diffusion by substituting equation \ref{Xa1 expression weak diffusion} into \ref{dXa/dz z=h definition},
\begin{equation}\label{final weak diffusion BC z=h}
\bigg[ \frac{\partial \chi_{A}}{\partial z} \bigg]_{h} = \frac{(J_A - J \chi_{A0})}{h \big( \frac{J}{3} - \frac{1}{Pe' E h} \big)}
\end{equation}
\subsection{Solution method and initial conditions}
\subsubsection{K\'{a}rm\'{a}n-Pohlhausen approximation}
We now apply the K\'arm\'an-Pohlhausen integral approximation whereby we integrate equations \ref{dimensionless mass}, \ref{dimensionless mom}, \ref{dimensionless eng}, and equation \ref{dimensionless conc 2} over $z$ from $0$ to $h$. Doing this removes any multiple variable differentials while retaining the inertia and advection terms in the momentum and energy balance equations. First, let us define the integrated forms of $f$ and $\Theta$ as,
\begin{equation}
f = \int_{0}^{h} u \,dz, \quad\quad\quad \Theta = \int_{0}^{h} T \,dz.
\label{F TH def}
\end{equation}
In order to be able to evaluate equation \ref{F TH def}, we now need to prescribe the forms of $u$, and $T$ as function of the vertical coordinate. To this end, we assume that each variable can be approximated by a polynomial of the form $c_1 + c_2 z + c_3 z^2$. By substituting the corresponding polynomials in equation \ref{F TH def} and applying the appropriate boundary conditions, it is possible to evaluate the polynomial constants and eventually derive the following expressions for $u$ and $T$,
\begin{equation}
u = \bigg(\frac{3 f}{h^2} - \frac{\partial \sigma}{\partial r} \frac{1}{2 \mu Ma} \bigg)z - \bigg(\frac{3 f}{2 h^3} - \frac{\partial \sigma}{\partial r} \frac{3}{4 h \mu Ma} \bigg)z^2
\label{closure approx u}
\end{equation}
\begin{equation}
T = T_w + \bigg(\frac{(J_A + \Lambda J_B)} {2 k} + \frac{3 \Theta} {h^2} -\frac{3 T_w} {h} \bigg)z + \bigg(- \frac{3(J_A + \Lambda J_B)} {4 h k} - \frac{3 \Theta} {2 h^3} + \frac{3 T_w} {2 h^2} \bigg)z^2
\label{closure approx T}
\end{equation}
Integration of the governing equations along with application of the boundary conditions defined in section \ref{interfacial boundary conditions} yields the following integrated forms of the mass, $r$-momentum, energy and concentration equation in the limit of weak diffusion,
\begin{equation}
\frac{\partial h}{\partial t} = - EJ - \frac{1}{r}\frac{\partial (rf)}{\partial r} - \frac{f}{r}
\label{height eq kp}
\end{equation}
\begin{equation}
\varepsilon Re \bigg(\frac{\partial f}{\partial t} + \frac{1}{r} \frac{\partial}{\partial r} \bigg(r \int_{0}^{h} u^2 \,dz \bigg) + u\vert_h EJ \bigg) = - h \frac{\partial p}{\partial r} + \bigg[ \mu \frac{\partial u}{\partial z} \bigg]_{0}^{h}
\label{r mom kp}
\end{equation}
\begin{equation}
\varepsilon Re Pr c_p \bigg( \frac{\partial \Theta}{\partial t} + \frac{1}{r} \frac{\partial}{\partial r} \bigg( r \int_{0}^{h} u T \,dz \bigg) + T\vert_h EJ \bigg) = \bigg[ k  \frac{\partial T}{\partial z} \bigg]_{0}^{h} 
\label{energy kp}
\end{equation}
\begin{equation}
\frac{\partial \chi_{A0}}{\partial t} + \frac{f}{h} \frac{\partial \chi_{A0}}{\partial r} = \frac{(J_A - J \chi_{A0})}{ Pe' h^2 \big( \frac{J}{3} - \frac{1}{Pe' E h} \big)}
\label{integrated conc weak diffusion}
\end{equation}
Note that in the above expressions, all terms containing $u$ and $T$ are evaluated using equations \ref{closure approx u} and \ref{closure approx T} and therefore we end up with expressions containing the unknown variables $f$ and $\Theta$ instead of $u$ and $T$.
\subsubsection{Precursor film and resulting boundary conditions}
As previously mentioned, we assume that the droplet is surrounded by a thin precursor film covering the heated substrate upon which it resides. In this region, the fluid is flat with zero mean curvature and sufficiently thin such that evaporation is suppressed by attractive van der Waals forces. We assume the mixture in the precursor region is at equilibrium concentration, $\chi_{A,\infty} = 0$, meaning that it consists solely of the LVC. Simplifying equation \ref{normal stress BC} subject to these conditions when $h = h_\infty$ yields the expression for precursor layer height:
\begin{equation}
h_\infty = \bigg( \frac{\mathcal{A} \delta} {\Lambda T \vert _h}\bigg)^{1/3}
\end{equation}
We now turn our attention to the boundary conditions at the bottom wall where the liquid meets the solid substrate ($z = 0$). Here, we impose conditions of no-penetration, no-slip, and constant temperature, such that:
\begin{equation}
\frac{\partial \chi_A}{\partial z} = 0, \quad \mathbf{u} = 0, \quad T = 1.
\end{equation}
Finally, we apply the following boundary conditions to the radial extremes of the domain ($r = 0$ and $r = r_\infty$),
\begin{equation}
\left. \begin{array}{l}
\displaystyle
\frac{\partial h}{\partial r}(0,t) = 0, \quad
f(0,t) = 0, \quad
\frac{\partial \Theta}{\partial r}(0,t) = 0, \quad
\frac{\partial \chi_A}{\partial r}(0,t) = 0; \\[16pt]
\displaystyle
h(r_\infty,t) = h_\infty, \quad
\frac{\partial h}{\partial r}(r_\infty,t) = 0, \quad
f(r_\infty,t) = 0, \quad
\Theta(r_\infty,0) = h_\infty, \quad
\chi_A(r_\infty,t) = 0.
\end{array} \right\}
\label{boundary conditions in r}
\end{equation}
\subsubsection{Penalty function}
Due to our modelling approach, the droplet is deposited onto a thin precursor film. This film is sufficiently thin so that van der Waals interactions in the liquid phase become the dominating force and hence suppress further evaporation in this precursor region. It is then logical to assume that the precursor layer consists solely of the LVC since any MVC will have evaporated before the film forms. When testing the model, we noticed that artificial behaviour can occur in the precursor film resulting from the added complexity of a second component. Diffusion of the MVC from the bulk droplet into the  into the precursor film is possible, as is condensation of MVC from the gas phase into the film region. To circumvent this problem, we incorporate a forcing-type penalty function ($\mathcal{P}$) with which we can control the composition of the precursor film. This ensures that the inert precursor region does not interfere with the evaporation of the droplet or induce any artificial behaviour.

The penalty function itself is applied to the advection-diffusion (concentration) equation and forces the precursor film to solely consist of the LVC, preventing any evaporation or condensation from occurring. It takes the form,
\begin{equation}
\mathcal{P} = \mathcal{M} \chi_{A0} \bigg( 1 - \tanh\bigg[\mathcal{B} \bigg( \frac{h}{h_\infty} -1 \bigg) \bigg] \bigg) = 0
\end{equation}
where $\mathcal{M}= 10^3$ is its magnitude and $\mathcal{B} = 5$. When $h > h_\infty$, as is the case in the bulk droplet, $\mathcal{P}$ is zero regardless of the value of concentration and so has no effect on the solution. The penalty function begins to influence the solution when droplet height approaches that of the precursor. If $h = h_\infty$, $\mathcal{P}$ tends towards $\mathcal{M}$. When applied to the conservation equation for concentration, $\chi_A$ is forced to zero, minimising $\mathcal{M}$ and ensuring in $\mathcal{P}$ is equal to zero once more. The physical effects of this restriction are twofold. First, it is ensured that there is no artificial condensation of the MVC into the precursor layer. Second, any diffusion of MVC from the bulk droplet to the precursor layer is arrested.
\subsubsection{Initial conditions}
\label{Initial conditions}
Within the droplet profile ($0 \leq r \leq 1$), the initial conditions are imposed such that:
\begin{equation}
h(r,0) = h_\infty + 1 - r^2, \quad
f(r,0) = 0, \quad
\Theta(r,0) = h(r,0) T_0, \quad
0 \le \chi_{A0,i} \le 1.
\label{initial conditions in 0 < r < 1}
\end{equation}
Here, $\chi_{A0,i}=\chi_A(r,0)$ is the initial uniform concentration within the droplet. Outside of the droplet in the precursor layer region ($r > 1$) we apply the following,
\begin{equation}
h(r,0) = h_\infty, \quad
f(r,0) = 0, \quad
\Theta(r,0) = h_\infty, \quad
\chi_{A0,i} = 0.
\label{initial conditions in r > 1}
\end{equation}
\subsubsection{Overview of solution procedure}
\begin{figure}
	\centering
	\includegraphics[width=0.8\textwidth]{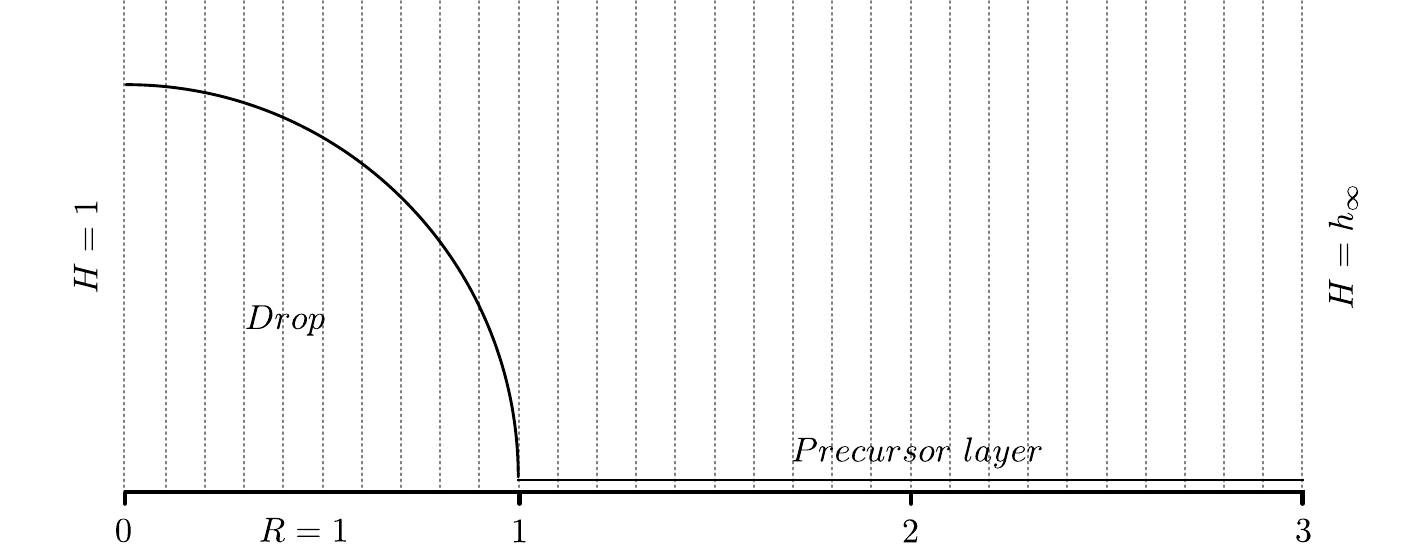}
	\caption{Illustration of the height, $h$, variable under initial conditions in a domain where $r_\infty = 3$. The one dimensional domain consists of equally spaced $N_r$ nodes, here, the vertical dotted lines represent every tenth node where the total number of nodes, $N_{r,tot} = 300$. The value of height is stored at every node point and is reconstructed to form the drop profile over the domain. The drop is initialised as a quarter circle in dimensionless space for $0 \le r \le 1$, with the precursor layer height, $h_\infty$, imposed for $r > 1$. Similar profiles along $r$ are used as initial conditions for the other variables---see section \ref{Initial conditions}.}
\label{discretised profile fig}
\end{figure}

From our definitions above, we have 7 unknown variables; $h$, $p$, $f$, $\Theta$, $J_A$, $J_B$, and $\chi_{A0}$ along with 7 independent equations. As a broad overview of the solution procedure, we begin with simplifying these equations by applying the Galerkin method of weighted residuals to obtain weak forms for each equation. Derivation and final forms of the weak equations are given in \citet{Williams2018}. The domain is discretised from $0$ to $r_\infty$ into a uniform mesh of $N_{r,tot}$ nodes (see figure \ref{discretised profile fig}) using the finite element method (FEM). Solutions are then obtained using a Newton-Raphson scheme with the simulation evolved forward in time using implicit Euler and an adaptive time step, $dt$. The time step is increased or decreased based on the largest residual error of the governing equations from the previous time step. Initial solutions are provided (via the initial conditions in section \ref{Initial conditions}) and progressively more accurate values iterated to over each time step. The iterative program is written in Fortran, making use of the linear algebra package LAPACK.
\section{Experimental methodology}
\subsection{Apparatus and experimental procedure}
\begin{figure}
	\centering
	\includegraphics[width=0.8\textwidth]{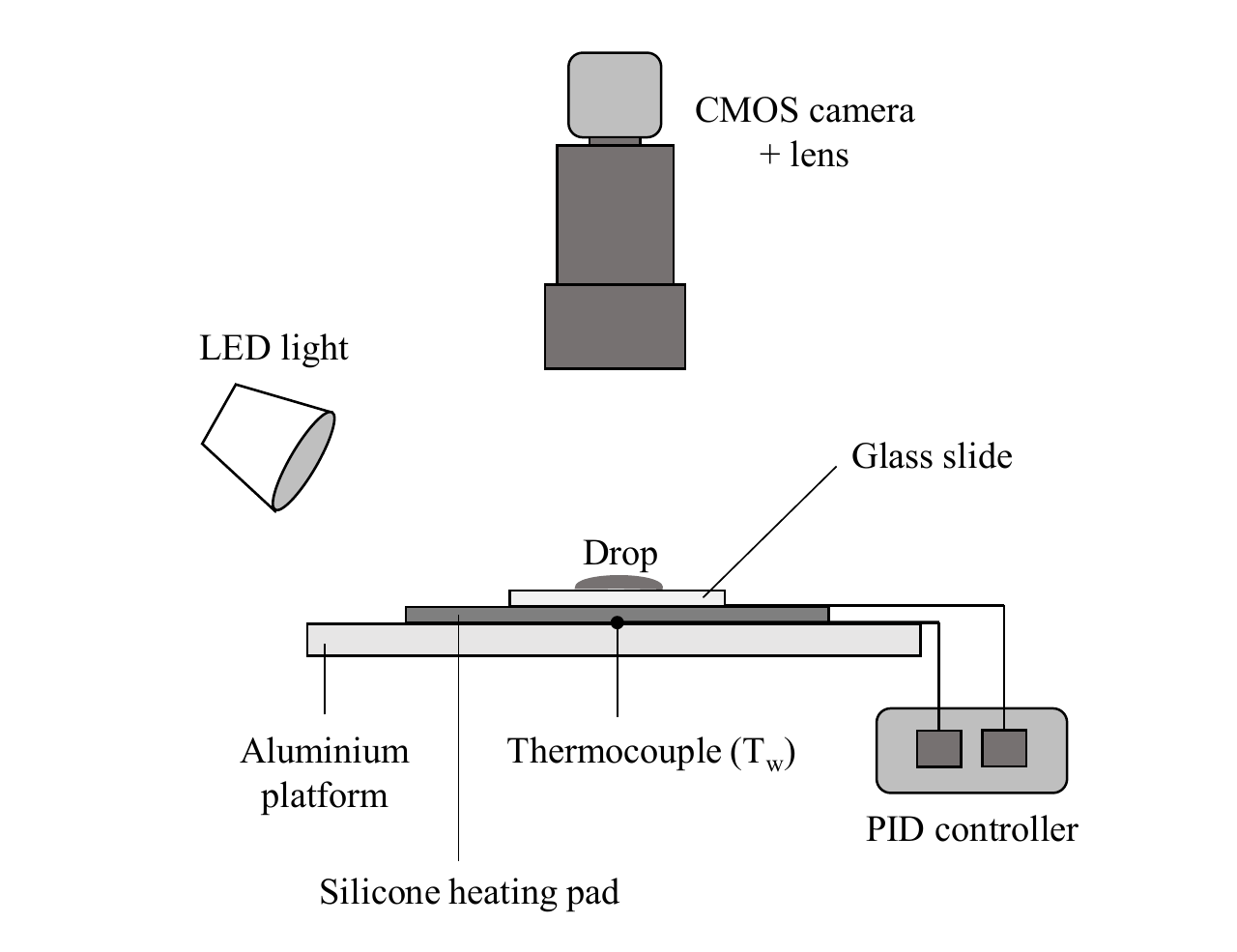} 
	\caption{Schematic diagram of the experimental apparatus.}
	\label{fig:spreading_drop_apparatus_diagram}
\end{figure}
A diagram of the experimental apparatus is shown in figure \ref{fig:spreading_drop_apparatus_diagram} which centres around a flexible silicone heating pad (Omega SRFR-4/5-P-230V) providing a heat flux of 0.775 \si{\watt\per\cm\squared}. This sits atop an aluminium mechanical scissor lift platform and is held in place with heavy duty white duct (Gorilla) tape. The temperature of the heater is controlled with a PID controller in a feedback loop; the controller maintains the desired set point measured by a thermocouple attached to the heating pad. The CMOS camera is held in place above the scissor lift platform using a laboratory stand and clamp with liberal amounts of duct tape securing it to the desk. The CMOS camera used is a Point Grey Research Flea3 (FL3-U3-13E4M) with a \SIrange{18}{108}{\milli\metre}/\SIrange{2.5}{16}{} Navigator Zoom \num{7000} zoom lens. The camera is connected to a PC via USB3 and is controlled through FlyCapture2 software. Optical recording is conducted at 60 fps. The droplet is illuminated from the side using a touch mounted on a large 3 prong clamp as the light source. To ensure a clear image is captured by the camera, Diall PVC repairing tape, possessing a smooth white surface, is layered on top of the duct tape. 

Borosilicate glass microscope slides (75 mm $\times$ 25 mm, 1 mm thick) manufactured by RC Components are used as the substrate. These are simply placed on top of the tape holding down the heating pad with the friction between the two materials sufficient to prevent movement. The glass slides consistently demonstrated a low equilibrium contact angle for all fluids tested. High wettability was verified by treating the slides with ``piranha'' solution---a volatile mixture of sulfuric acid and hydrogen peroxide. Piranha solution is a strong oxidiser and so removes organic matter whilst additionally hydroxylating the surface. The droplets are deposited on the substrate manually using a microliter syringe (Hamilton 701N \SI{10}{\micro\litre}) with reading increments of \SI{0.2}{\micro\litre}.

We consider ethanol-water mixture droplets of initial volume \SI[separate-uncertainty = true]{1.0(2)}{\micro\litre}. Mixtures ranging from \SI{11}{\wtpercent} to \SI{50}{\wtpercent} initial ethanol concentration are considered at three substrate temperatures ($T_w$); \SI{30}{\celsius}, \SI{50}{\celsius} and \SI{70}{\celsius}. Solutions are prepared in \SI{25}{\milli\litre} volumes and stored in \SI{25}{\milli\metre} diameter jars. Separate syringes of volume \SI[separate-uncertainty = true]{2.50(5)}{\milli\litre} were used to collect samples of each pure component for mixing. The mixing volumes of each fluid as well as the initial ethanol concentrations investigated are given in table \ref{mixing volume table}. Once the solutions are prepared, evaporation of the mixtures was kept to a minimum by covering the mouth of the jar with a plastic paraffin film (Parafilm); this allowed the seal to be retained with the lid removed. A sample was taken by piercing the film with the micro-syringe, leaving only a small hole and suppressing unwanted evaporation as much as possible. The lid was returned after obtaining each sample. For each mixture concentration deposited on each substrate temperature, a minimum of five experimental runs were conducted to ensure the results are replicable. 

\begin{figure}
	\centering
	\includegraphics[width=0.7\textwidth]{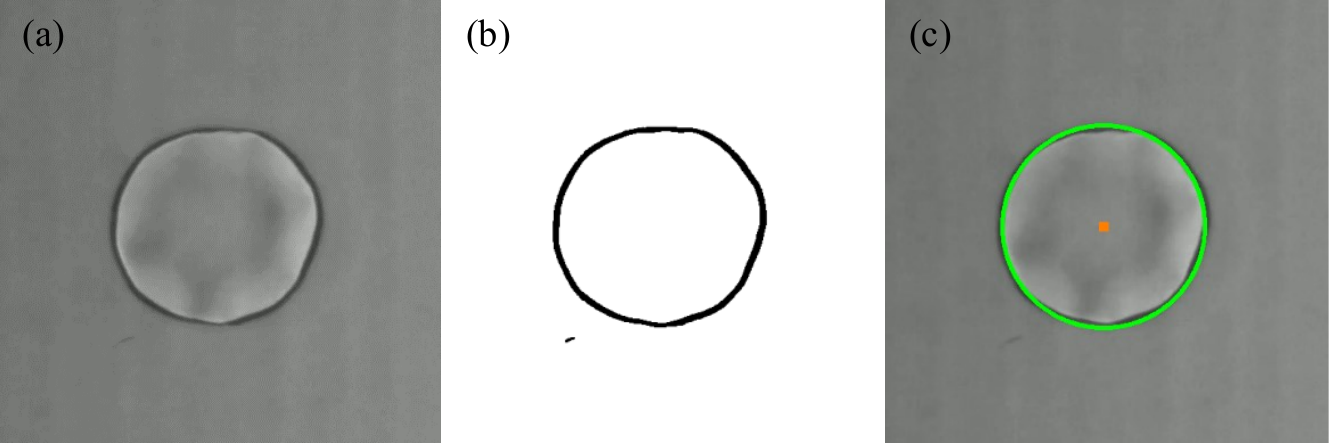} 
	\caption{Top-down view snapshots of a \SI{1}{\micro\litre} ethanol-water droplet comprising \SI{25}{\wtpercent} initial ethanol deposited on a \SI{70}{\celsius} substrate at $t =$ \SI{0.6}{\second}. (a) shows the original greyscale image captured by the camera, (b) shows the binary image after passing through imaging filters, and (c) shows the best-fit circle (green) to the contact line (black) along with the corresponding  centre point (orange) overlaid on (a).}
	\label{image processing}
\end{figure}
The results are processed by tracking the droplets radius over time, both the initial spreading followed by contact line recession as evaporation takes over. The radius is tracked frame-by-frame using an in-house algorithm written in python, making use of NumPy and OpenCV libraries. The basic overview is to convert each frame to a high contrast image using in-built OpenCV image processing tools and then detect the circular shape of the droplet using the OpenCV Hough Circles Transform. Image processing begins by removing noise from the greyscale images captured by the camera by passing through the GuassianBlur and medianBlur filters. After this, the sharp edges of the image corresponding to the contact line are detected using the adaptive threshold filter and converted to a binary black and white image using the binary threshold filter. The Hough Circles Transform is applied to this image, which then determines the best fit circle to the circular-shaped droplet outline and calculates the corresponding centre point and radius. To set the scale, a circular black sticker of diameter \SI{0.8}{\mm} is affixed to a sample glass slide. With the scale set, the expanding and contracting radius of the droplet as it spreads and recedes is measured directly. A clear limitation of this method is that the droplet must be close to circular to obtain meaningful results. In our case, this is already a requirement since we are comparing to a 1D axisymmetric model where the droplet is perfectly circular. Contact line radius against time for each droplet can then be plotted. The spreading and retraction rates are obtained by analysing the radius-time graphs in the common logarithmic domain using R statistical software \citep{R2013} made available under the GNU General Public Licence. This method allows linear fits along with breakpoints to be determined in a statistically significant and consistent manner.
\subsection{Errors and uncertainty}
We briefly discuss the sources of error in the experiment, some more difficult to quantify than others. Table \ref{mixing volume table} gives the error in measuring the volumes of ethanol and water when preparing the binary mixtures for storage. These are typically low and based on the reading error of the syringes used to prepare the mixtures. The final volume of droplet deposited on the substrate is subject to larger error. Each \SI{1}{\micro\litre} droplet is deposited using a microsyringe with reading increments of \SI{0.2}{\micro\litre}. Assuming a reading error of \SI{\pm0.1}{\micro\litre} yields a \SI{10}{\percent} relative error in the deposited volume. In addition to this, we noticed that there was often a small amount of liquid residue left on the tip of the syringe after deposition. As such, the relative error in the deposited volume is likely to be larger than \SI{10}{\percent}, with a \SI{20}{\percent} relative error in the volume deposited being a worst case prediction. The uncertainly from the PID feedback loop can be assumed as \SI{\pm 1}{\kelvin}. However, with the heater and thermocouple buried beneath an insulating plastic tape along with inherently low thermal conductivity of the glass substrate, it is likely that the surface the droplet is deposited onto will be slightly cooler than the displayed value by the controller. 

Considering imaging errors, a clear droplet image is captured by the angled light source casting a shadow around the contact line. This causes the contact line to appear thicker than in reality. In addition, the formation of a ridge at the contact line in droplets with higher initial ethanol concentration causes this region to appear thicker still. Contact line instabilities also arise in ethanol rich droplets, making accurate resolution even more difficult. Measuring the pixel width of the droplet at its thickest point in the final images provides a reasonable estimate of this error. Our radius detection method relies on the idealistic assumption that droplets are always perfectly circular throughout spreading and recession. In the absence of perfectly consistent curvature around the whole circumference, the algorithm will fit a circle that best fits the largest portion of the droplet circumference. This results in fluctuation of the radius measurement as the algorithm searches for the optimum curvature. The best estimation of this uncertainty comes from the standard error of the linear fit determined by R.

\begin{table}
\begin{center}
\def~{\hphantom{0}}
\setlength{\tabcolsep}{8pt}
\begin{tabular}{S[table-format=2.2(1),separate-uncertainty = true,table-align-uncertainty=false]
                S[table-format=2.2(1),separate-uncertainty = true,table-align-uncertainty=false]
                S[table-format=2.1(1),separate-uncertainty = true,table-align-uncertainty=false]
                S[table-format=2.1(1),separate-uncertainty = true,table-align-uncertainty=false]
                }
{Ethanol (\si{\milli\litre})}	&{Water (\si{\milli\litre})}	&{Initial ethanol \si{\volpercent}} 	&{Initial ethanol	\si{\wtpercent}}\\[3pt]
0.00			        &25.00(50)	        	    &0.0			        &0.0\\
3.50(10)		        &21.50(45)			        &14.0(7)	        	&11.4(6)\\
7.50(15)		        &17.50(35)			        &30.0(12)		        &25.3(10)\\
14.00(30)		        &11.00(25)  	            &56.0(30)               &50.0(27)\\
\end{tabular}
\caption{Mixing volumes of ethanol and water used to prepare the mixtures and the corresponding initial volume and weight percentages of ethanol.}
\label{mixing volume table}
\end{center}
\end{table}
To minimise this error for each run, we took several measures to maximise even spreading of the droplets. These include ensuring a completely level surface, the selection of small droplet volumes, and the gentle deposition of the droplets from the microsyringe. Another limitation worth mentioning is that, particularly for higher concentrations of ethanol, droplets do not dry out in a circular shape meaning the exact point of dry out cannot be measured by our algorithm. Rather, we rely on the visual disappearance of the droplet from the original video footage for this.
\section{Experimental findings}\label{experimental findings}
\subsection{Typical evaporation process}
As previously mentioned, we consider only droplets of pure water and water-ethanol mixtures consisting of \SI{11}{\wtpercent}, \SI{25}{\wtpercent}, and \SI{50}{\wtpercent} initial ethanol at substrate temperatures of \SI{30}{\celsius}, \SI{50}{\celsius}, and \SI{70}{\celsius}. In order to maximise the evaporation rate for comparison with our simulations, we restrict our investigations into the effect of concentration variation for a substrate at temperature $T_w =$ \SI{70}{\celsius} only, while effects of temperature variation are restricted to the most volatile binary mixture---\SI{50}{\wtpercent} initial ethanol. Higher ethanol concentrations, extending to pure ethanol are not included due to difficulties in capturing a sharp contact line using our imaging method.

After a droplet is deposited carefully with the microsyringe, the typical evaporation process for all concentrations and temperatures can be split into two main stages: a rapid spreading stage followed by a slower retraction stage. These stages are to be expected with wetting droplet and has been observed extensively in the literature \citep{Semenov2014}. The length of each stage depends on the droplet composition and substrate temperature. Additionally, for lower volatility cases, a third stationary phase can appear between spreading and retraction whereby the droplet remains at maximum radius for a time before retraction begins. Such behaviour is also expected for lower volatility liquids \citep{Cachile2002a} and is observed in our modelling results for low evaporation numbers---see, for example, figure \ref{compare E Xa = 050}.

Immediately after depositions, the droplets spread to their maximum radius. The very initial stages are dominated by inertial spreading, similar to pure and other binary mixture droplets \citep{Winkels2012,Mamalis2018}. Table \ref{spreading const rmax T = 70C table} gives the spreading coefficients, $n$ (where $R \propto t^n$), for each linear regime and their corresponding breakpoints in time, $b$, to the next linear regime. The maximum radius achieved by each drop is given by $r_{max}$. A visual representation of table \ref{spreading const rmax T = 70C table} is shown in figure \ref{exp srates constT graph}. Here, the experimentally measured radii are plotted against time on a log-log scale with the best fit lines ($n$) for each regime and transition breakpoints ($b$) between regimes also drawn. In the case of pure water (first column of table \ref{spreading const rmax T = 70C table} and figure \ref{exp srates constT graph}(a)), the inertial spreading exponent, $n_1$, is \num[separate-uncertainty = true]{0.36(7)}. $n_1$ increases when ethanol is added to the mixture, as seen in the remaining three columns of table \ref{spreading const rmax T = 70C table} and figures \ref{exp srates constT graph}(b), (c), and (d), meaning inertial spreading proceeds at a faster rate for higher initial ethanol concentration. After the inertial phase, spreading rate then decreases to a viscous regime, characterised by spreading exponents close to Tanner's law in the case of pure water and higher for binary ethanol-water compositions. After maximum radius is reached, droplets possessing lower volatilities and those on cooler substrates remain stationary for a period of time before retraction. In the case of binary droplets, retraction tends to happen in two stages; an initial rapid retraction followed by a slower contact line recession at later times. We now examine these processes in more detail for a \SI{25}{\wtpercent} and \SI{50}{\wtpercent} ethanol-water droplet on a \SI{70}{\celsius} substrate.
\begin{table}
\begin{center}
\def~{\hphantom{0}}
\setlength{\tabcolsep}{8pt}
\begin{tabular}{lS[table-format=2.2(1),separate-uncertainty = true,table-align-uncertainty=true]
                S[table-format=2.2(1),separate-uncertainty = true,table-align-uncertainty=true]
                S[table-format=2.2(1),separate-uncertainty = true,table-align-uncertainty=true]
                S[table-format=2.2(1),separate-uncertainty = true,table-align-uncertainty=true]
                }
\multicolumn{1}{c}{} &\multicolumn{4}{c}{$\chi_{A0,i}$}\\[3pt]
			                    &{0.00}         &{0.11}       &{0.25}     &{0.50} \\[6pt]
$n_1$		                    &0.36(7)        &0.74(16)       &1.61(11)       &3.66(33)    \\
$b_1$(\si{\second})             &0.65(17)       &0.63(20)       &0.87(14)       &0.24(1)    \\[4pt]
$n_2$		                    &0.23(3)        &0.54(13)       &1.15(45)       &1.36(15)    \\
$b_2$(\si{\second})	            &1.29(10)       &1.30(17)       &1.20(12)       &0.65(3)    \\[4pt]
$n_3$		                    &0.09(4)        &0.30(11)       &0.45(37)       &0.59(6)    \\
$b_3$(\si{\second})             &2.14(14)       &2.13(14)       &1.63(9)        &1.68(4)    \\[4pt]
$n_4$		                    &0.00           &0.02(4)        &-0.34(12)      &-0.03(6)   \\
$b_4$(\si{\second})             &7.49(59)       &4.87(8)        &2.73(04)       &           \\[4pt]
$n_5$		                    &-0.23(2)       &-0.71(27)      &-2.06(24)      &			\\
$b_5$(\si{\second})	            &21.87(3)       &5.87(4)        &3.69(4)        &			\\[4pt]
$n_6$		                    &-0.78(4)	    &-2.31(32)      &0.07(30)       &			\\
$b_6$(\si{\second})	            &33.16(1)       &5.77(3)        &4.47(6)        &			\\[4pt]
$n_7$		                    &-2.74(16)      &-0.37(3)       &-1.34(14)      &			\\
$b_7$(\si{\second})	            &               &14.87(9)       &6.81(19)       &			\\[4pt]
$n_8$		                    &               &-0.93(9)       &-0.86(6)       &			\\
$b_8$(\si{\second})	            &			    &20.33(5)       &14.42(7)       &			\\[4pt]
$n_9$		                    &			    &-2.14(16)      &-1.98(14)      &			\\[8pt]
$r_{max}$(\si{\milli\meter})    &2.33(11)       &3.01(14)       &4.47(12)       &5.35(30)   \\
\end{tabular}
\caption{Experimentally measured spreading exponents, $n$, corresponding breakpoints in time, $b$, and maximum radii, $r_{max}$ for ethanol-water sessile droplets for increasing initial concentrations of ethanol, $\chi_{A0,i}$, at substrate temperature $T_w=$ \SI{70}{\celsius}.} \label{spreading const rmax T = 70C table}
\end{center}
\end{table}
\subsection{25 wt.\% ethanol-water droplet}
\begin{figure}
	\centering
	\includegraphics[width=0.90\textwidth]{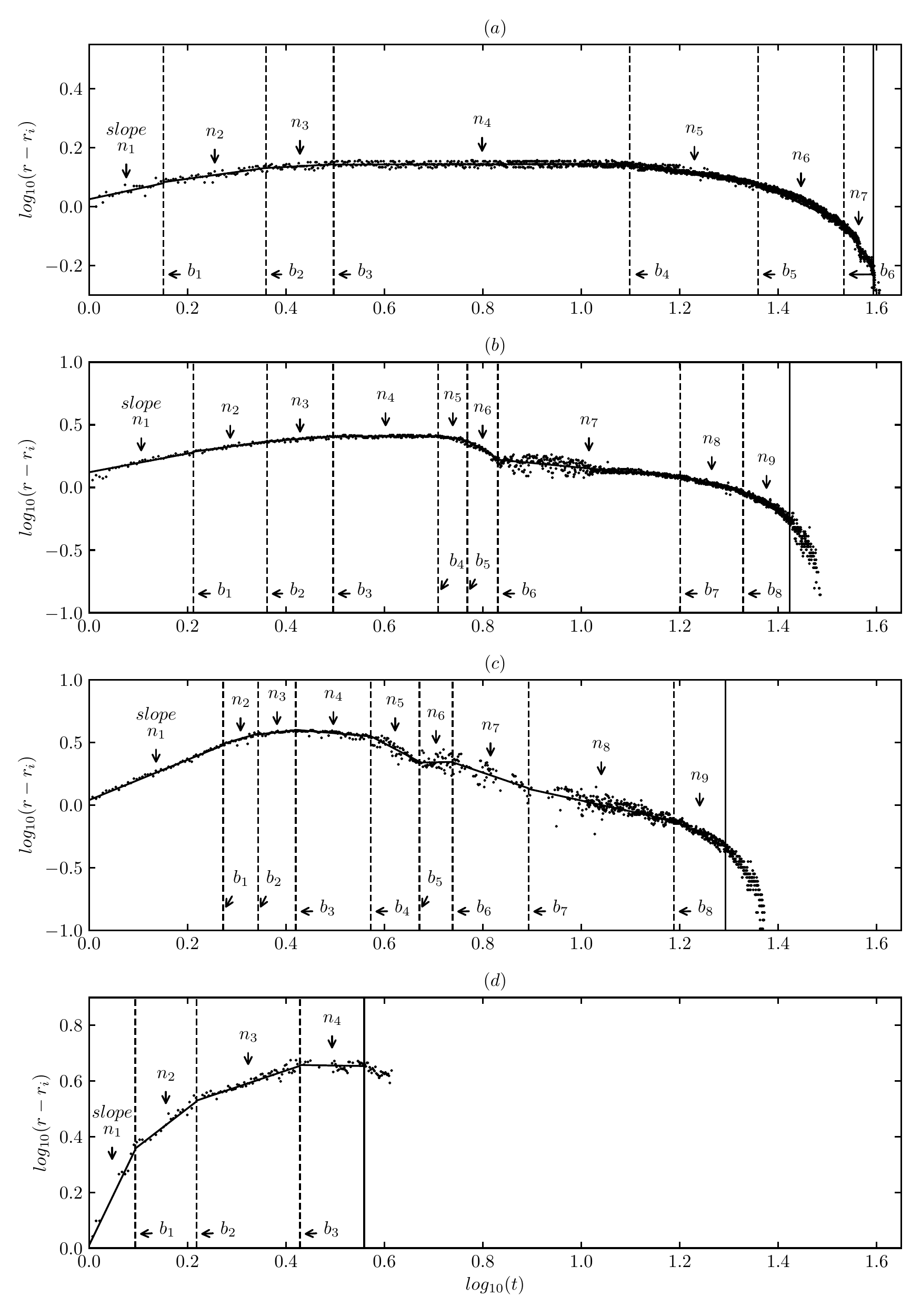}
	\caption{Experimentally measured droplet radii against time for droplets deposited on a substrate at $T_w=$ \SI{70}{\celsius}. Droplet radius $r$ is normalised by the first recorded radius after deposition, $r_i$ and plotted in the logarithmic space along with time after deposition. Spreading rates, $n$, for each regime are shown as best fit lines and the breakpoints, $b$, signifying transition to the next linear regime drawn as vertical dashed lines. Initial ethanol concentration ($\chi_{A0,i}$) for each plot is as follows; (a) $\chi_{A0,i} = 0.00$ (pure water) , (b) $\chi_{A0,i} = 0.11$, (c) $\chi_{A0,i} = 0.25$, and (d) $\chi_{A0,i} = 0.50$. See table \ref{spreading const rmax T = 70C table} for the corresponding numeric values of $n$ and $b$ for each $\chi_{A0,i}$.}
	\label{exp srates constT graph}
\end{figure}

\begin{figure}
	\centering
	\includegraphics[width=0.9\textwidth]{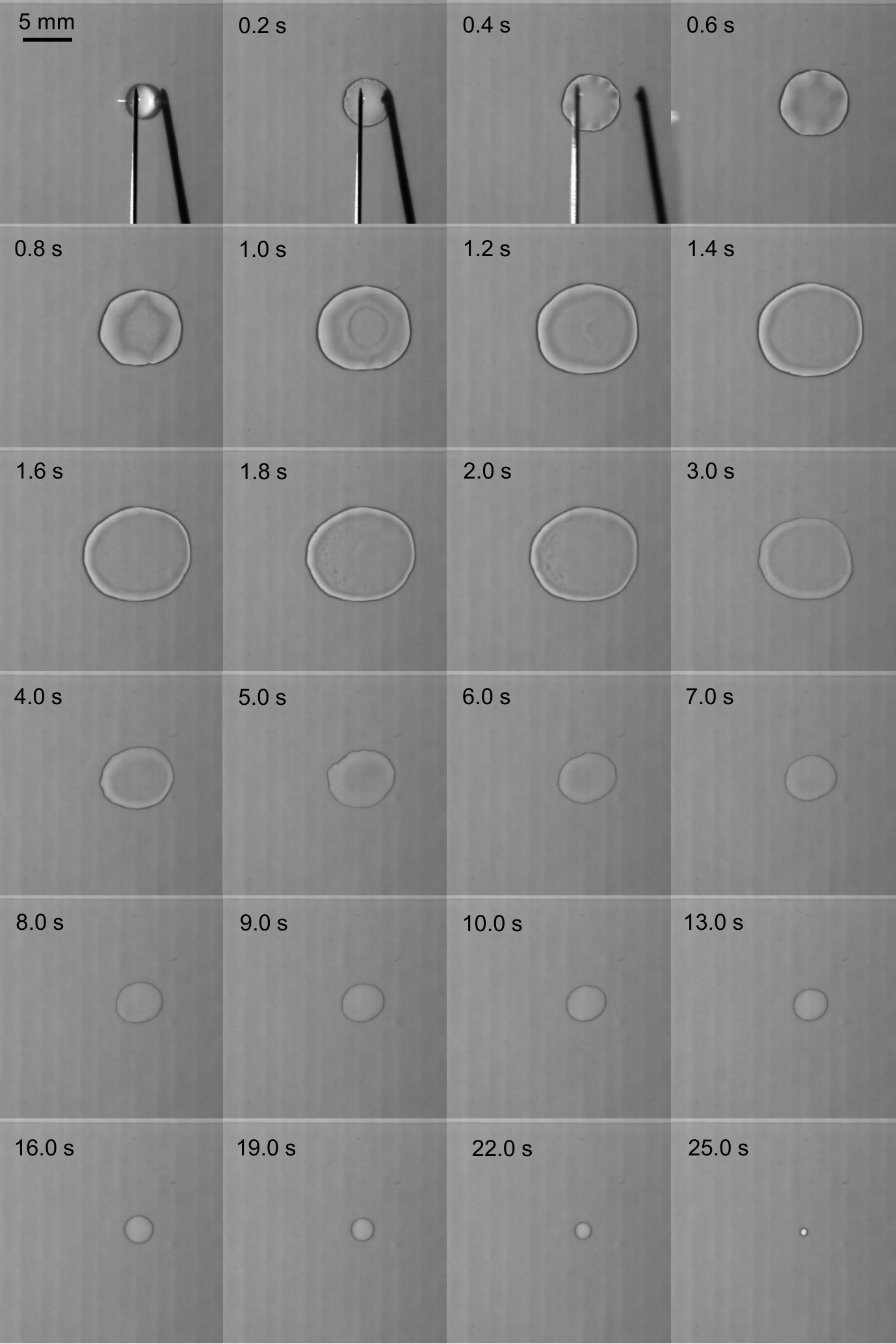} 
	\caption{Top-down view snapshots of a \SI{1}{\micro\litre} ethanol-water droplet comprising \SI{25}{\wtpercent} initial ethanol deposited on a \SI{70}{\celsius} substrate.}
	\label{exp drop snapshots 25}
\end{figure}

Figure \ref{exp drop snapshots 25} presents snapshots taken with the CMOS camera over the lifetime of a \SI{25}{\wtpercent} ethanol-water droplet on a \SI{70}{\celsius} substrate. The third column of table \ref{spreading const rmax T = 70C table} gives the spreading exponents and their transition points in time for this concentration with a visual representation given in figure \ref{exp srates constT graph}(c) . After deposition at $t = 0\,$\si{\second}, the droplet begins to spread rapidly with $n_1 = 1.61 \pm 0.11$ up until $t = 0.87 \pm 0.14\,$\si{\second}, considered to be firmly within the inertial regime. Faint interface ripples appear near the contact line at  $t = 0.4\,$\si{\second}, subsequently dying down by $t = 0.8\,$\si{\second} as the spreading rate slows slightly to $n_2 = 1.15 \pm 0.45$. The lighter rim near the droplet edge indicates a thicker area of liquid near the contact line, presumably formed from strong currents pulling the fluid outwards. The droplet continues to spread until $t \approx 2.0\,$\si{\second} while at the same time the light rim decreases in thickness. A maximum droplet radius of $r = 4.47 \pm 0.12\,$\si{\milli\metre} is reached. The droplet then proceeds to recede in two main regimes. A period of rapid recession comes first with an exponent, $n_5 = -2.06 \pm 0.24$, terminating at $t = 3.69 \pm 0.04\,$\si{\second}. The second regime is slower and characterised by an exponent of $n_8 = -0.86 \pm 0.06$. Our simulations indicate that the first rapid recession is owing to the sudden reversal of surface tension gradient as ethanol becomes sufficiently depleted within the droplet. The droplet then continues to evaporate and recede until dry-out at $t \approx 25.0\,$\si{\second}.
\subsection{50 wt.\% ethanol-water droplet}\label{50 ethanol-water droplet}
\begin{figure}
	\centering
	\includegraphics[width=0.9\textwidth]{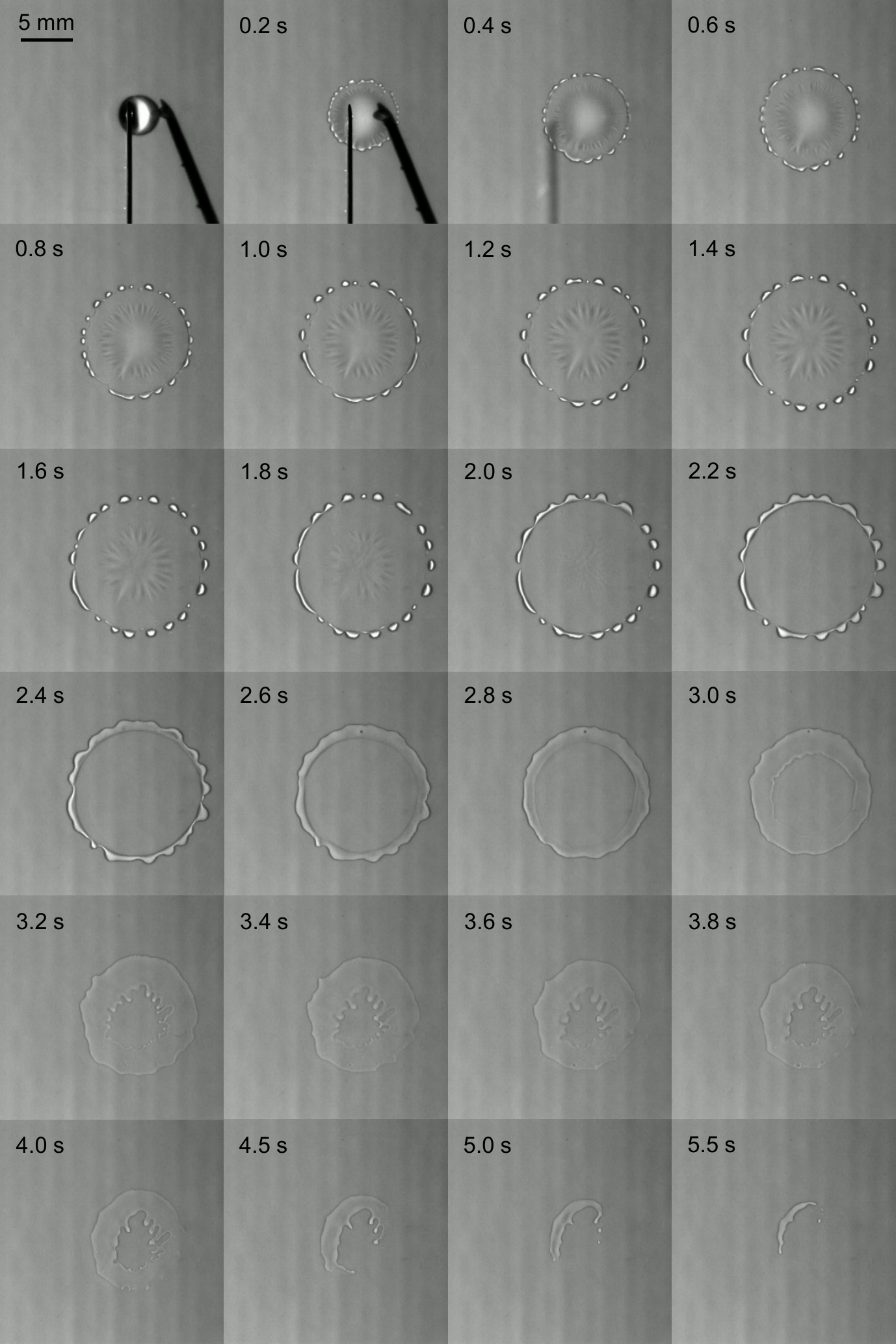} 
	\caption{Top-down view snapshots of a \SI{1}{\micro\litre} ethanol-water droplet comprising \SI{50}{\wtpercent} initial ethanol deposited on a \SI{70}{\celsius} substrate.}
	\label{exp drop snapshots 50}
\end{figure}

Upon increasing the initial concentration of ethanol from \SI{25}{\wtpercent} to \SI{50}{\wtpercent}, radically different behaviour emerges. Figure \ref{exp drop snapshots 50} shows camera stills taken over the droplet lifetime and the corresponding spreading exponents are given in the fourth column of table \ref{spreading const rmax T = 70C table} and shown visually by figure \ref{exp srates constT graph}(d). It is immediately clear when comparing with the lower concentration droplet in figure \ref{exp drop snapshots 25} that the initial spreading rate when $\chi_{A,i} = 0.50$ is noticeably faster. Beginning at $n_1 = 3.66 \pm 0.33$ until $t_1 = 0.24 \pm 0.01\,$\si{\second} and continuing at the slightly reduced rate of $n_2 = 1.36 \pm 0.15$ until $t_2 = 0.65 \pm 0.03\,$\si{\second}. Spreading then proceeds at a rate of $n_3 = 0.59 \pm 0.06$ until the maximum radius of $5.35 \pm 0.30\,$\si{\milli\metre} is reached at $t_3 = 1.68 \pm 0.04\,$\si{\second}. From $t = 0.2\,$\si{\second} in figure \ref{exp drop snapshots 50}, two distinct instabilities can be seen forming in the droplet. The first is a contact line instability whereby the contact line breaks up into fingers that grow with time. The second instability appears to occur over the interface, equidistant between the droplet centre and contact line. It takes the form of spoke-like patterns arranged radially around the droplet centre, similar to those observed by \citet{Semenov2014}. 

The fingering instability at the contact line resembles the ``octopi'' instability observed by \citet{Mouat2020} and \citet{Gotkis2006} and is similar to the droplet ejection phenomena seen by \citet{Keiser2017} in ethanol-water droplets and \citet{Mouat2020} in isopropanol-water droplets. Since the emergence of both instabilities only occurs at high initial ethanol concentrations, the clear indication is that they arise due to solutal Marangoni stresses. As the droplet is initially deposited as a spherical cap, evaporation will be particularly strongest at the contact line---as we have predicted with our model. Preferential evaporation of ethanol at the contact line results in high ethanol concentration within the droplet, causing a large surface tension gradient between the apex and contact line and therefore driving rapid spreading. It is this rapid spreading that causes the fingering contact line instability. The spoke-line patterns on the interface appear to be resulting from the strong outward flow within the droplet towards the contact line. 

As time proceeds from $t = 0.2\,$\si{\second} to $t = 1.8\,$\si{\second}, figure \ref{exp drop snapshots 50} clearly shows the contact line fingers growing in volume while the number stays constant at 21--24 fingers. The thicker fingers appear white to the camera compared to the thinner droplet interior. Our theoretical model seems to predict this phenomena in 1D by the formation of a thicker ridge of liquid ahead of the contact line---see figure \ref{Xa=050 h profile snapshots}a. By $t = 2.0\,$\si{\second}, finger growth ceases and the radial interface patterns decay to leave a smooth interface. The droplet then begins to retract, although this could not be recorded by our detection algorithm due to the contact line not being sharp enough after passing through imaging filters. This sudden retraction, resulting from the reversal of the surface tension gradient as ethanol is depleted, causes the fingering patters to also decay as the contact line is drawn inwards. At this point, the droplet is likely to be constituted entirely of water. At around  $t = 3.2\,$\si{\second}, the droplet centre appears to dry out as it recedes, resulting in the formation of a second, inner contact line. We are now essentially left with a ring of liquid similar to that observed by \citet{Guena2007b}. This is also confirmed by our numerical model that predicts dry-out of the interior before the contact line ridge. With the formation of the inner contact line comes a third instability, emerging as inward facing fingers forming along the circumference of the inner contact line.
\subsection{Variation in concentration}
\begin{figure}
	\centering
	\includegraphics[width=\textwidth]{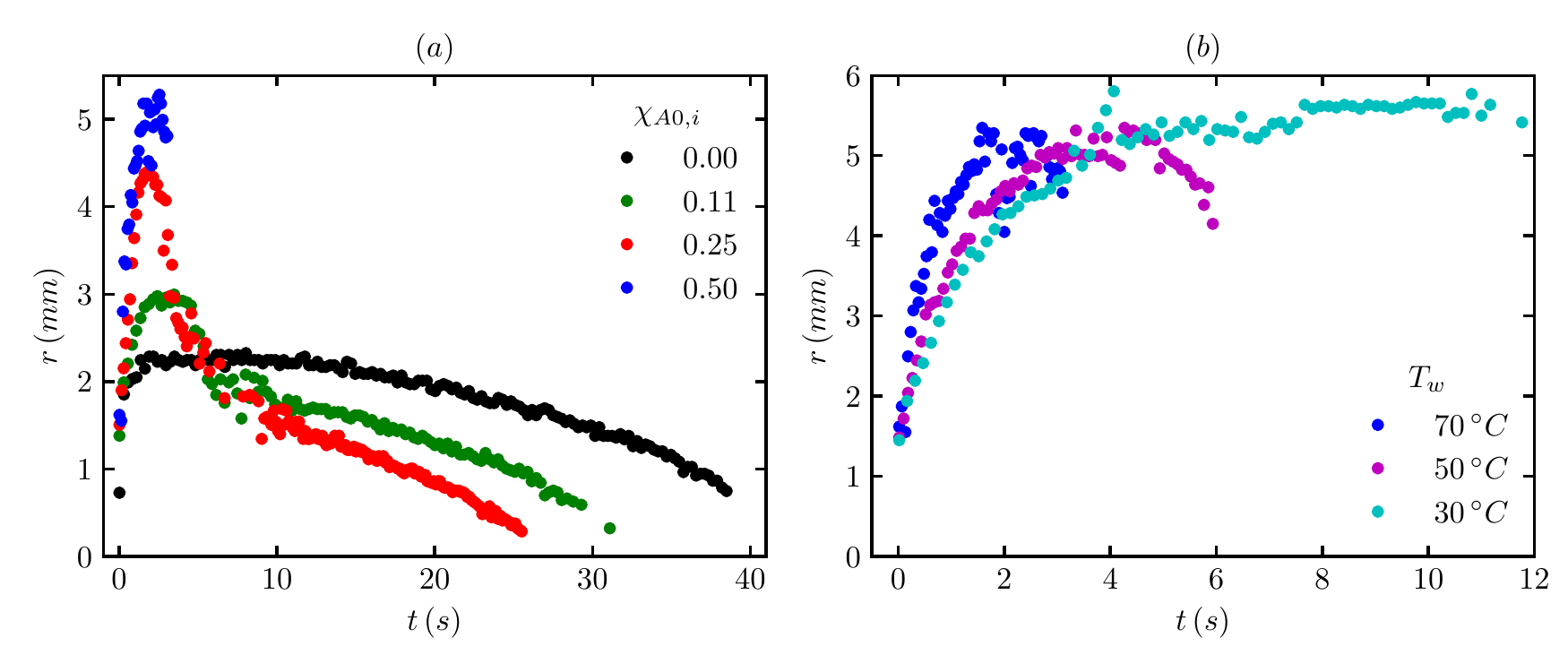} 
	\caption{Droplet radius versus time for (a) constant substrate temperature of \SI{70}{\celsius} for initial ethanol concentrations of \SIrange{0.00}{0.50}{\wtpercent}, and (b) initial ethanol concentration of \SI{50}{\wtpercent} for substrate temperatures of \si{30}, \si{50}, and \SI{70}{\celsius}. The error in the measurement of radius are $\pm 0.41$ mm (at $30 \celsius$), $\pm 0.24$ mm (at $50 \celsius$) and $\pm 0.30$ mm (at $70 \celsius$).}
	\label{exp radii plot}
\end{figure}

Figure \ref{exp radii plot}(a) plots the droplet radii measured by our detection algorithm for $\chi_{A,i} =$ \num{0.00}, \num{0.11}, \num{0.25}, and \num{0.50} versus time for $T_w =$ \SI{70}{\celsius}. This clearly illustrates the increased spreading (both rate and maximum radius) exhibited as initial ethanol concentration is increased. As expected, droplet lifetime decreases with increasing ethanol concentration, owing part to increased mixture volatility and part to a larger effective area for evaporation as spreading increases. Table \ref{spreading const rmax T = 70C table} also gives the maximum radii, $r_{max}$, achieved by the droplets in these plots. Compared to the \SI{1}{\micro\litre} pure water droplet, where $r_{max} = 2.33 \pm 0.11\,$\si{\milli\metre}, maximum radius is increased by \SI{29}{\percent} for a $\chi_{A,i} = 0.11$ droplet of the same volume and then by \SI{92}{\percent} and \SI{130}{\percent} for droplets of $\chi_{A,i} = 0.25$ and $\chi_{A,i} = 0.50$ respectively. The rapid recession regimes are also seen clearly for $\chi_{A,i} = 0.11$ and $\chi_{A,i} = 0.25$ in figure \ref{exp radii plot}(a), whereas recession is slow and steady for pure water. 
\subsection{Variation in temperature}\label{Variation in temperature}
We consider briefly the effects of varying the substrate temperature, $T_w$, restricting ourselves to only the most volatile ethanol-water mixture, $\chi_{A0,i} = 0.50$. Figure \ref{exp radii plot}(b) plots radius over time for $T_w =$ \SI{30}{\celsius}, \SI{50}{\celsius}, and \SI{70}{\celsius}. As we would expect, lower $T_w$ results in prolonged droplet lifetimes with the mixture volatility decreasing with temperature. Lower temperature droplets are therefore able to spread for longer times, achieving a larger $r_{max}$. It is also clear from figure \ref{exp radii plot}(b) that although droplets spread further overall, the rate of spreading is reduced as the substrate temperature is lowered. The spreading exponents for each regime along with maximum radii are given in table \ref{spreading const rmax conc Xa = 050 table}. As substrate temperature is increased, the spreading exponent for each regime increases while the corresponding break point in time signifying transition to the next regime occurs earlier. This is likely due to the more rapid development of a concentration gradient when the droplet touches the substrate as ethanol evaporates more vigorously at the higher temperatures. \citet{Mamalis2018} also saw an increase in the spreading exponents with substrate temperature in their experiments with self-rewetting droplets. Additionally, when the temperature is increased, the number of fingers produced at the contact line (see figure \ref{exp drop snapshots 50} and section \ref{50 ethanol-water droplet} for a detailed discussion of this instability) also increases, with approximately 18 seen at $T_w =$ \SI{30}{\celsius}, 20 at $T_w =$ \SI{50}{\celsius} and 21--24 seen at $T_w =$ \SI{70}{\celsius}. The finger length, which we define as the distance from the apparent contact line of the bulk droplet to the apex of the extended finger, also increases with substrate temperature as higher evaporation rate drives the instability. A similar trend was seen by \citet{Sefiane2010}, where the wavenumber of interfacial HTWs increased with increasing substrate temperature for FC-72 droplets, albeit driven by a different phenomenon viz. thermocapillary instabilities in a pure fluid. 
\begin{table}
\begin{center}
\def~{\hphantom{0}}
\setlength{\tabcolsep}{8pt}
\begin{tabular}{lS[table-format=2.2(1),separate-uncertainty = true,table-align-uncertainty=true]
                S[table-format=2.2(1),separate-uncertainty = true,table-align-uncertainty=true]
                S[table-format=2.2(1),separate-uncertainty = true,table-align-uncertainty=true]}
\multicolumn{1}{c}{} &\multicolumn{3}{c}{$T_w$}\\[3pt]
		        &{\SI{30}{\celsius}} &{\SI{50}{\celsius}}	&{\SI{70}{\celsius}}\\[6pt]
$n_1$		                    &1.29(10)       &2.01(15)       &3.66(33)	\\
$b_1$(\si{\second})             &0.96(1)        &0.50(1)		&0.24(1)	\\[4pt]
$n_2$		                    &0.64(6)        &0.82(6)		&1.36(15)	\\
$b_2$(\si{\second})             &2.15(4)        &1.53(3) 		&0.65(3)	\\[4pt]
$n_3$		                    &0.39(4)        &0.4(4)		    &0.59(6)	\\
$b_3$(\si{\second})	            &4.51(14)       &3.06(3) 		&1.68(4)	\\[4pt]
$n_4$		                    &-0.01(1)       &-0.13(5)		&-0.03(6)	\\[8pt]
$r_{max}$(\si{\milli\meter})    &5.85(41)       &5.4(24)        &5.35(30)    \\
\end{tabular}
\caption{Spreading coefficients, $n$, corresponding breakpoints in time, $b$, and maximum radii, $r_{max}$, at initial ethanol concentration of $\chi_{A0,i} = 0.50$ for increasing substrate temperatures at \SI{30}{\celsius}, \SI{50}{\celsius}, and \SI{70}{\celsius}.}
\label{spreading const rmax conc Xa = 050 table}
\end{center}
\end{table}
\section{Numerical results}
\subsection{The pure fluid limit}
\subsubsection{Validation}
\begin{figure}
	\centering
	\includegraphics[width=0.75\textwidth]{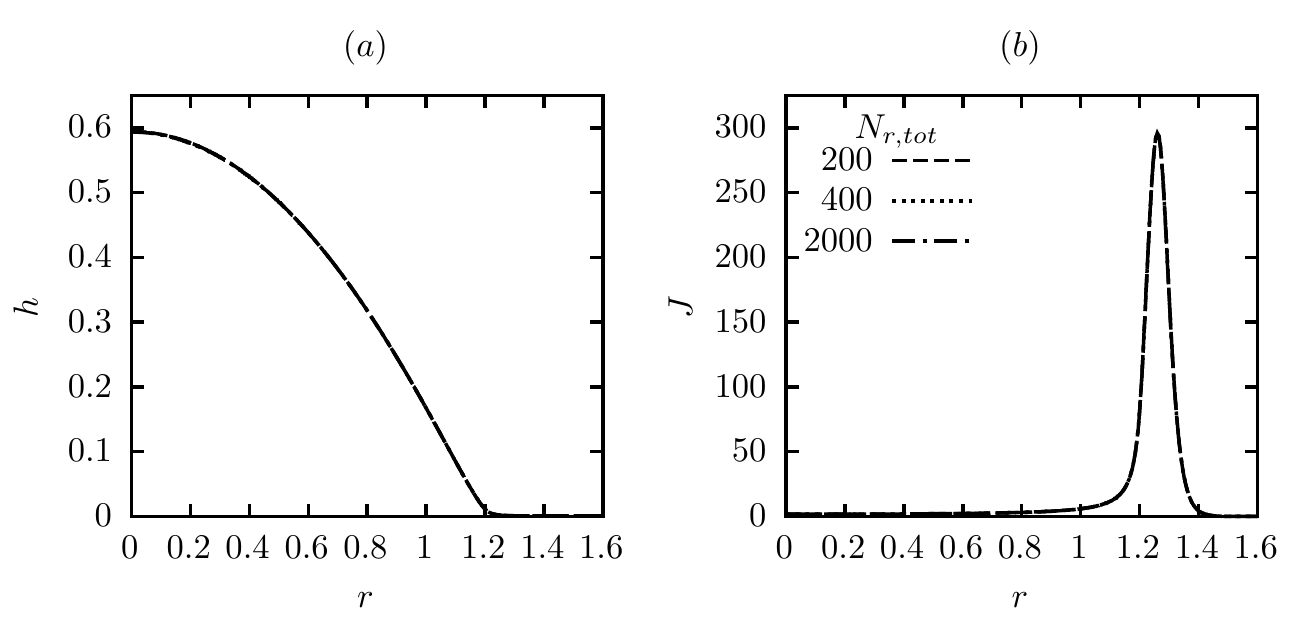}
	\caption{Snapshots of (a) interface profile, $h$, (b) total evaporative flux, $J$, of a droplet with $\chi_{A0,i} = 0.5$ with the remaining dimensionless properties are given in \ref{ethanol-water dimensionless base parameters table}. All property ratios set to unity, resembling a pure mixture. The domain length, $r_{\infty}$, is 2 and the number of nodes ($N_{r,tot}$) in  is increased from 200 to 400 to 2000, demonstrating grid independence of the solution.}
	\label{equal prop h J refine fig}
\end{figure}

\begin{figure}
\centering
	\includegraphics[width=0.75\textwidth]{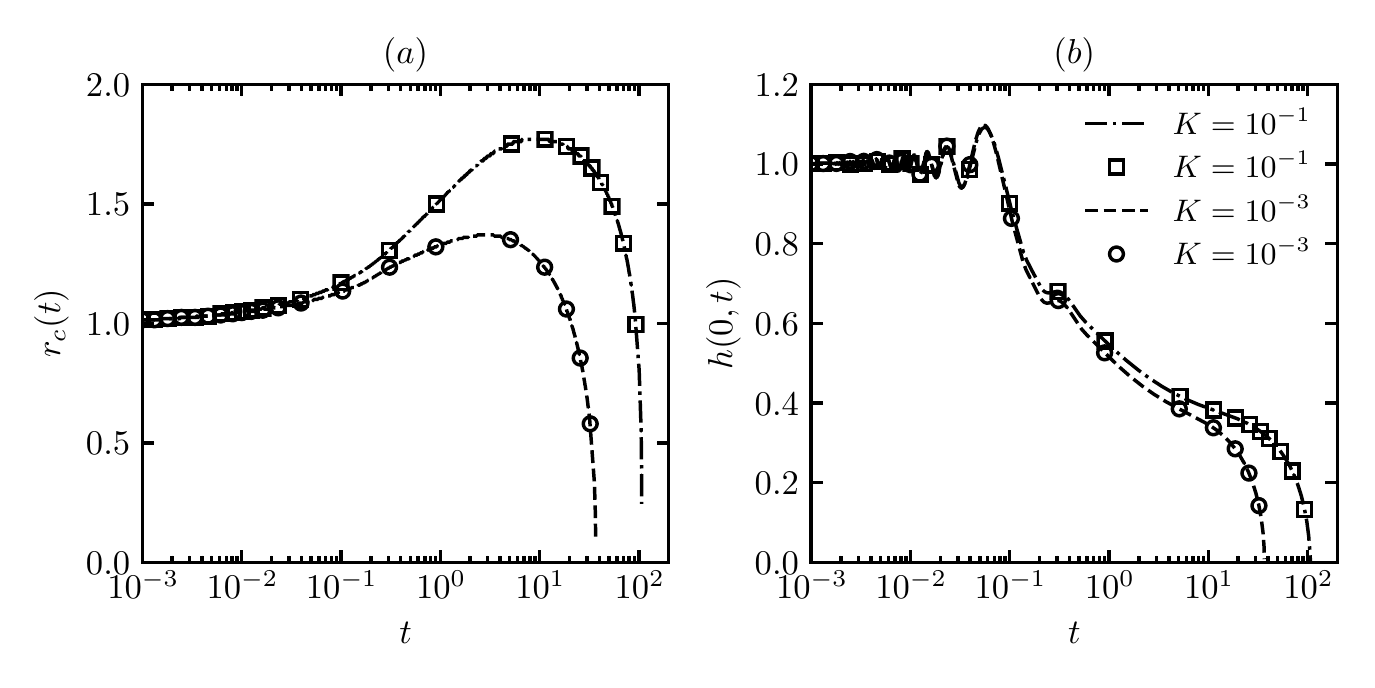}
	\caption{Comparison of the current model (dashed lines) with the pure fluid model of \citet{Karapetsas2010} (shown by symbols) for $K=10^{-3}$ and 0.1, (a) shows position of the contact line over time and (b) the height of the droplet apex over time; $\chi_{A0,i} = 0.5$ with all property ratios set to unity, resembling a pure mixture. The remaining dimensionless properties are  $\varepsilon = 0.2$, $Re = 5$, $Pr = 10$, $Ma = 10^{-2}$, $Pe = 25$, $E = 10^{-3}$, $\delta = 10^{-5}$, and $\mathcal{A} = 10^{-4}$.}
\label{equal prop r h fig}
\end{figure}
Returning now to our one-sided model defined in section \ref{model definition}, we first validate our model against the pure fluid model by \citet{Karapetsas2010} on which ours is based. To approximate a single component mixture, all property ratios are set to unity and the initial mass fraction, $\chi_{A0,i}$ to $0.5$. This effectively mimics a pure fluid---an equal mixture of two identical components. A domain length of $r_{\infty} = 2$ is used with total number of elements, $N_{r,tot} = 200$. Grid convergence is demonstrated in \ref{equal prop h J refine fig} where the total number of nodes is refined to $N_{r,tot} = 400$ and $N_{r,tot} = 2000$, with the same independent solutions obtained using all meshes.

Figure \ref{equal prop r h fig} shows the contact line position, $r_c$, and apex height, $h(0,t)$, for two values of the Knudsen number; $K = 10^{-3}$ and $K = 0.1$. As expected, the results from our pseudo-single component model agree well with the solutions of \citet{Karapetsas2010} (symbols overlaying the dashed lines). Oscillations at the apex are observed at early times when $t < 10^{-1}$ due to inertia at $Re=5$. Calculated from dimensional properties, $K \approx 10^{-3}$, however, the evaporation rate can be controlled by increasing $K$ which effectively decreases the heat transfer rate and evaporation across the interface. Figure \ref{equal prop r h fig} shows that increasing $K$ to $0.1$ prolongs the droplet life time resulting in a longer spreading time and maximum droplet radius before evaporation takes over and the contact line begins to recede.
\subsubsection{Pure water droplet}
\begin{table}
\begin{center}
\def~{\hphantom{0}}
\setlength{\tabcolsep}{10pt}
\begin{tabular}{ll|ll|ll}
$\varepsilon$   &\num{0.2}      &$\delta$       &\num{1e-5}     &$k_R$      &\num{1.00}	\\
$Re$            &\num{0}        &$\mathcal{A}$  &\num{1e-4}     &$\mu_R$	&\num{0.84} \\
$Pr$			&\num{16.1}		&$Pe$ 			&\num{5}        &$c_{p,R}$	&\num{1.74} \\
$Ma$			&\num{1.64e-1}  &$\sigma_R$     &\num{3.20}		&$M_R$ 		&\num{0.39} \\
$E$				&\num{2.66e-4}  &$\gamma_R$ 	&\num{1.81}     &$\Lambda$	&\num{1.00}	\\
$K$				&\num{8.85e-4}  &$\alpha$		&\num{0.40}     &$\chi_{A0,i}$ & 0-0.75  \\
\end{tabular}
\caption{Typical dimensionless base parameters for an ethanol-water mixture}
\label{ethanol-water dimensionless base parameters table}
\end{center}
\end{table}
We now introduce the parameters used in modelling an ethanol-water droplet. We begin by assuming a temperature difference between the substrate and air, $\Delta \hat{T}$, of \SI{45}{\celsius}. All droplets have an initial volume of \SI{1}{\micro\litre} and an initial aspect ratio of \num{0.2}. Dimensionless numbers and property ratios are calculated from the physical properties of each component given in table \ref{ethanol-water table}, and listed in table \ref{ethanol-water dimensionless base parameters table}. The droplets we consider are assumed to be small and very thin, meaning, surface tension is the dominating force. Thus, we focus on the Stokes flow limit and we also set $Pe=5$ such that $\varepsilon^2 Pe \approx 1$, as required by our theory. This will also help suppression of the interfacial oscillations seen in figure \ref{equal prop r h fig} for most cases. The P\'{e}clet number indicates the rate of mass diffusion in the droplet; high numbers indicate slow diffusive component transport. Mass transport is intimately tied to the rate of evaporation, something that is relatively fast in our one-sided model due to the assumption of a phase-transition limited evaporation over a diffusion limited approach.

The parameters, $\mathcal{A}$ and $\delta$ are set to $10^{-4}$ and $10^{-5}$ respectively and we assume both components have equal latent heats ($\Lambda = 1$). This sets the precursor thickness ($h_\infty$) to $10^{-3}$, corresponding to 1/1000th of the initial apex height of the droplet. The precursor layer in our model will be thicker than in experiments which are wildly regarded to be in the submicron range around \SI{100}{\angstrom} \citep{DeGennes1985, Bonn2009}. If we assume the \SI{1}{\micro\litre} droplets from our experiment are initially deposited (however momentarily) as a perfect spherical cap, the initial apex height will be approximately \SI{3/4}{\mm}. A precursor thickness of \SI{100}{\angstrom} will therefore be around 1/75000th of the initial apex height, making the precursor layer in our model almost 2 orders of magnitude larger. We are forced into the compromise of 
$h_\infty = 10^{-3}$ because an overly thin precursor layer results a very large disjoining pressure in our model, causing the problem to become numerically stiff and convergence hard to achieve. Decreasing either $\mathcal{A}$ or $\delta$ individually by an order of magnitude (resulting in $h_\infty \approx$ \num{5e-4}) has a very minor effect on the solution. Lastly, for simplicity, we also assume a uniform thermal conductivity throughout the droplet, meaning $k_R = 1$. The remaining dimensionless number and property ratios are left as the directly calculated quantities from the liquid component properties given in table\ref{ethanol-water table}.

\begin{figure}
	\centering
	\includegraphics[width=0.5\textwidth]{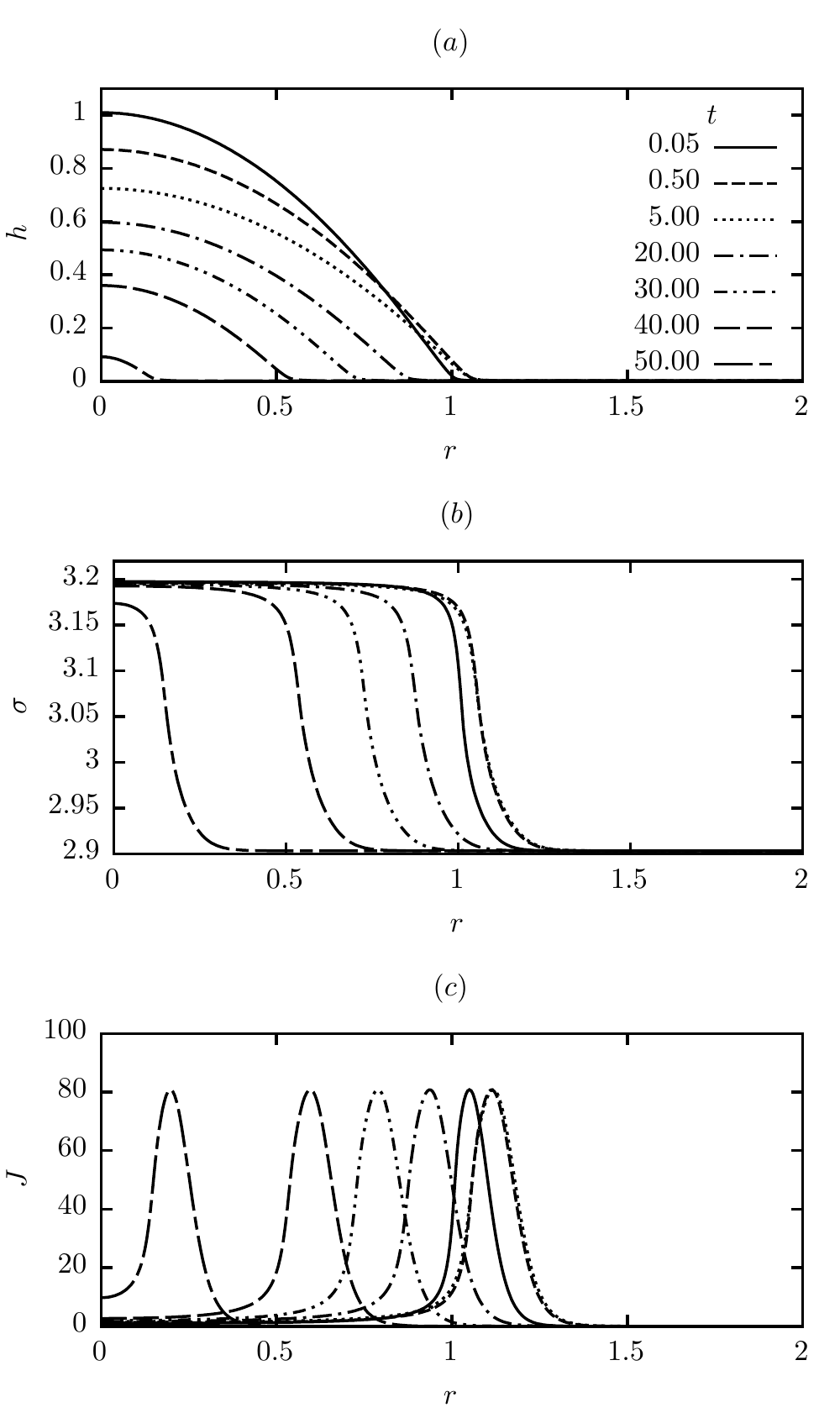}
	\caption{Snapshots of (a) interface profile, $h$, (b) surface tension, $\sigma$, and (c) total evaporative flux, $J$, of a pure water droplet over its lifetime. Dimensionless parameters are those given in table \ref{ethanol-water dimensionless base parameters table} with $\chi_{A0,i} = 0$.}
	\label{pure water h profile snapshots}
\end{figure}
Before considering a binary ethanol-water droplet, we first study the spreading and evaporation behaviour of a pure water droplet to serve as a reference case. A pure water droplet corresponds to the dimensionless properties in table \ref{ethanol-water dimensionless base parameters table}, with $\chi_{A0,i} = 0$. Figure \ref{pure water h profile snapshots} details the evolution of the interface profile, surface tension, and total evaporative flux along $r$ via snapshots in time as the droplet evaporates. The interface begins with a scaled dimensionless height and radius of \num{1}. At early times, the droplet spreads outwards as the forces at the contact line come into balance. By $t = 5$, evaporation takes over and the contact line slowly recedes with the droplet retaining a spherical cap shape over the remaining lifetime until dry-out at $t \approx 50$. The heated substrate causes the droplet to always be warmest at the contact line due to the reduced thickness of the liquid. It is evident that throughout the droplet lifetime, maximum evaporation occurs at the warm contact line---see figure \ref{pure water h profile snapshots}(c), where the vapour pressure is highest. The minimum liquid temperature is always located at the droplet apex. In the absence of solutal Marangoni effects, this is also the location of highest surface tension. Figure \ref{pure water h profile snapshots}(b) shows that a positive surface tension gradient between the contact line and apex is maintained throughout the droplet lifetime. Thermal Marangoni stresses therefore drive the liquid from the contact line towards the apex, limiting spreading in the early stages and causing the spherical cap to be retained as evaporation takes over and the contact line recedes. This behaviour is in line with the findings in other similar theoretical and experimental works \citep{Ehrhard1991,Ehrhard1993}, and with the mechanisms described by \citet{Deegan2000} and \citet{Hu2006}.
\subsection{Binary mixture droplet behaviour}
\begin{figure}
	\centering
	\includegraphics[width=0.75\textwidth]{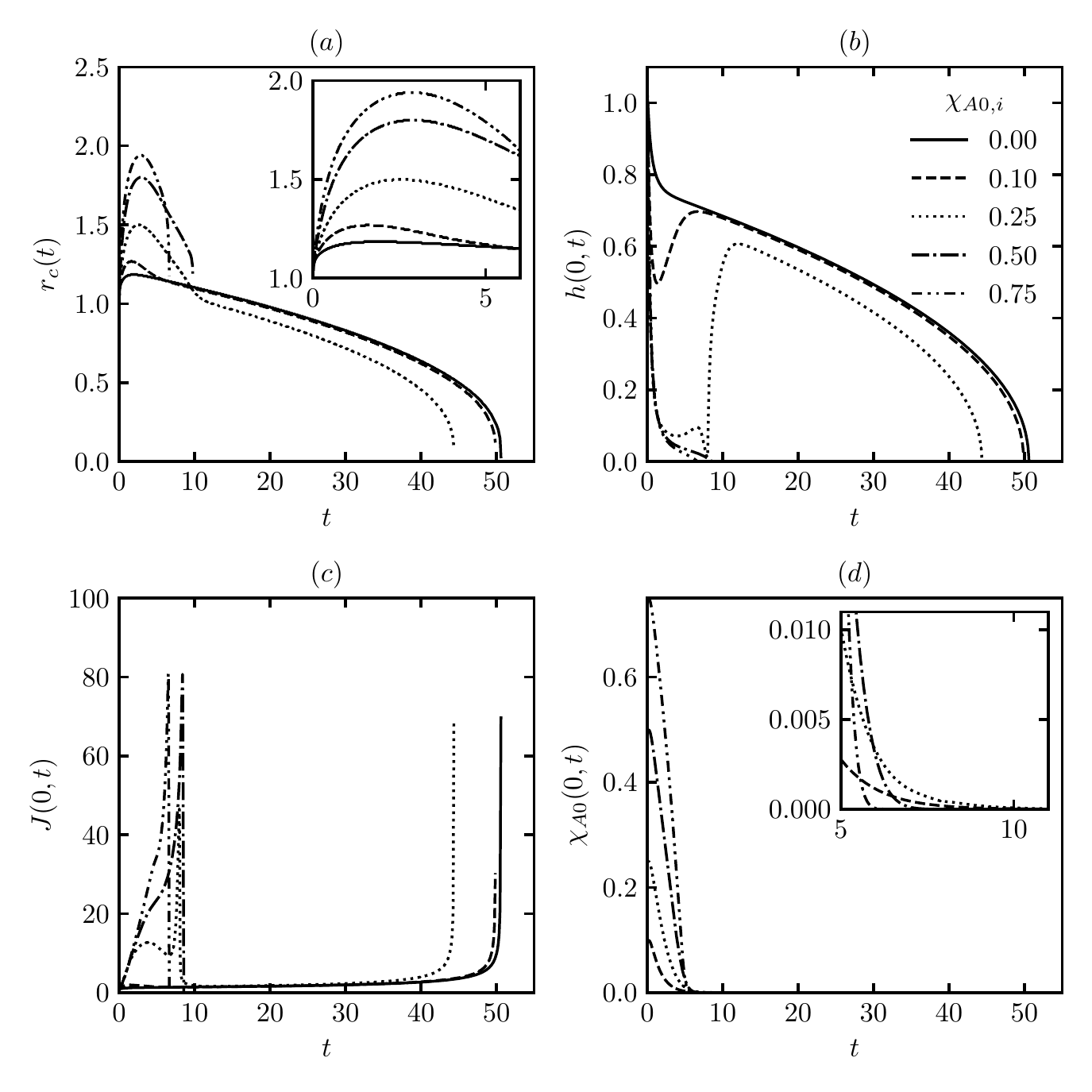}
	\caption{Profiles of (a) contact line position, (b) apex height, (c) apex mass flux, and (d) apex mass fraction throughout droplet lifetime for varying initial mass fraction of MVC, $\chi_{A0,i}$. Dimensionless parameters are given in table \ref{ethanol-water dimensionless base parameters table} with only $\chi_{A0,i}$ altered in each dataset.}
	\label{ethwater Xa t graph}
\end{figure}
We now gradually increase the initial mass fraction of ethanol ($\chi_{A0,i}$) in the droplet and examine the effects this has on the spreading behaviour and total lifetime. The parameters used are again those in table \ref{ethanol-water dimensionless base parameters table}. Specifically, we look at five cases: $\chi_{A0,i} = 0.00,\,0.10,\,0.25,\,0.50,\,0.75$. Figure \ref{ethwater Xa t graph} shows the position of the contact line, apex height along with the total evaporative flux and mass fraction of ethanol at the apex versus time. Beginning by again considering a pure water droplet, figure \ref{ethwater Xa t graph}(a) shows that pure water sees a modest initial spreading followed by a steady recession. After the initial stages, the height also decreases steadily---see figure \ref{ethwater Xa t graph}(b)---and evaporation from the apex is modest until the final stages before dry-out---figure \ref{ethwater Xa t graph}(c). Introducing ethanol into the droplet, we see that increasing $\chi_{A0,i}$ enhances the droplet spreading and increases the maximum position of the contact line. In all cases, the enhanced spreading is accompanied with a rapid droplet in apex height. Droplet lifetime is reduced as $\chi_{A0,i}$ increases owing both to the increased volatility of the mixture and the decreased droplet thickness due to enhanced spreading.

For $\chi_{A0,i} = 0.10$, we see that once a maximum radius is reached, the droplet begins to retract, accompanied by a regain in apex height to a position similar to the pure water droplet. Closer inspection of figure \ref{ethwater Xa t graph}(d) reveals that contact line retraction coincides with depletion of $\chi_{A0}$ at the apex, and hence in the rest of the droplet. A similar behaviour is displayed by $\chi_{A0,i} = 0.25$, with a greater initial spreading and maximum radius followed by a smaller retracted radius due to the larger proportion of evaporated ethanol leaving less droplet mass once depleted. Beyond this, with droplets constituting mainly water, evaporation then proceeds in the same way as the pure water droplet until dry-out. 

\subsubsection{Mechanisms governing contact line motion}
In both of these cases, enhanced spreading is driven by the preferential evaporation of ethanol from the contact line. This leaves an ethanol depleted (water rich) region at the contact line with higher surface tension than the bulk droplet. Induced by solutal Marangoni stresses, liquid flows towards the freely moving contact line, causing it to spread further outwards. Spreading continues until ethanol is depleted at which point solutal Marangoni stresses are eliminated. With the absence of ethanol, there is no longer any solutal Marangoni stress and the surface tension gradient is reversed with only thermal Marangoni stress present in the pure liquid. Surface tension now becomes highest in the coldest region of the droplet. On our heated substrate this corresponds to the thickest area of liquid, in these cases the apex. Flow is now directed away from the contact line towards the apex, driven now by thermal Marangoni stresses. The further the droplet has spread and deformed from its equilibrium shape, the further it must contract to regain this profile. With greater spreading at higher initial ethanol concentrations, this explains the rapid recession of the contact line and increase in height for $\chi_{A0,i} = 0.25$ over $\chi_{A0,i} = 0.10$ (see figure \ref{pure water h profile snapshots}a). It is clear that thermal and solutal Marangoni stresses are in competition with solutal effects dominating the initial stages and thermal effects the latter. We will look at these in more detail to follow.

\begin{figure}
	\centering
	\includegraphics[width=0.65\textwidth]{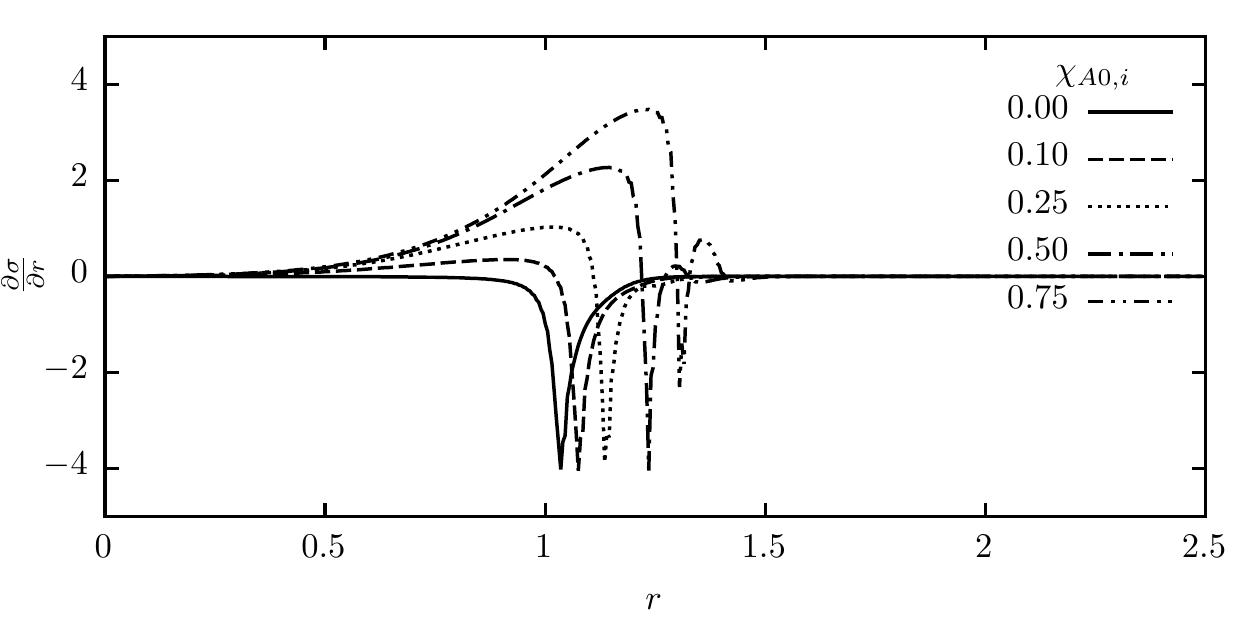}
	\caption{Rate of change of surface tension along $r$ for all initial ethanol concentrations considered at $t = 0.25$. Unless otherwise stated, dimensionless parameters are those given in table \ref{ethanol-water dimensionless base parameters table}.}
	\label{Sx all t=0.25}
\end{figure}
In the concentrations discussed previously, a significant amount of water remains after ethanol depletion, causing retraction and return to spherical cap shape. With higher initial ethanol, this is not the case and droplets remain in a flattened shape throughout their lifetime. Contact line recession in these binary mixtures is caused by both the inward driven Marangoni flow and mass loss from the droplet as it evaporates. Increasing initial ethanol from  $\chi_{A0,i} = 0.50$ to $\chi_{A0,i} = 0.75$, the droplet spreads by a greater amount---reaching a larger maximum radius. This is explained by the increased maximum surface tension gradient between the apex and the contact line for larger $\chi_{A0,i}$. Figure \ref{Sx all t=0.25} shows the change of surface tension along $r$ at the early time of $t = 0.25$ for the full range of concentrations considered. A positive surface tension gradient between the apex and contact line is clearly seen to increase with $\chi_{A0,i}$. A greater maximum spreading radius also results in a thinner droplet which is subject to higher temperatures and hence more rapid evaporation rate. Figure \ref{ethwater Xa t graph}(c) shows that there is always higher evaporative flux from the apex for higher initial ethanol concentration. This is due in part to the increased proportion of volatile ethanol but also to the decreased thickness causing a warmer interface and greater evaporation rate for any given mixture as well as the larger radius leading to an increased effective interfacial area for evaporation.

\begin{figure}
	\centering
	\includegraphics[width=0.5\textwidth]{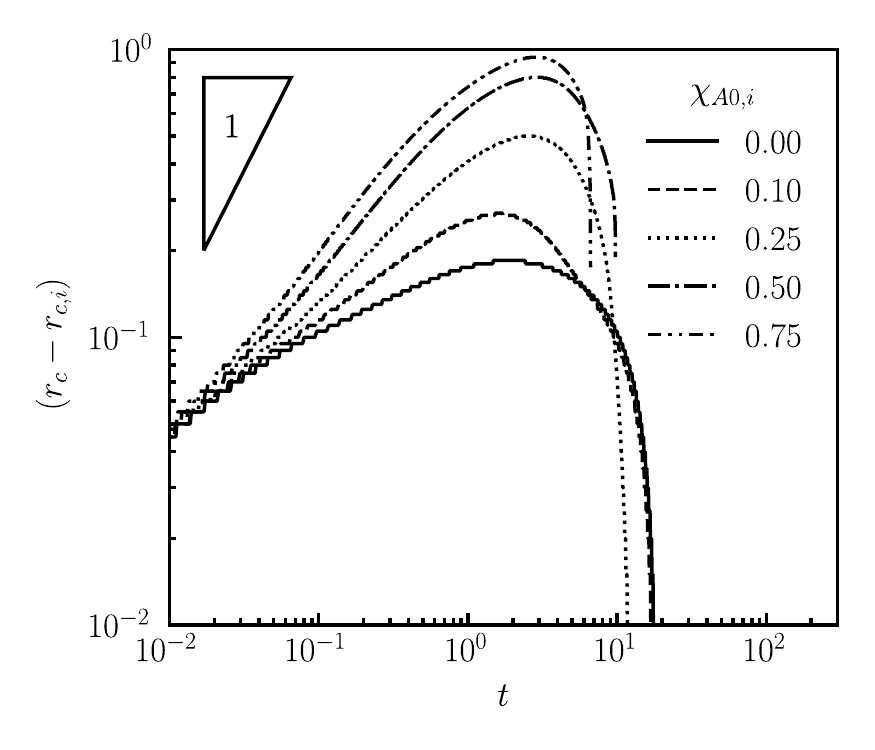}
	\caption{Contact line position versus time on a logarithmic scale for increasing initial ethanol concentrations. Corresponding spreading coefficients and breakpoints in time are shown in table \ref{modelling spreading coeffs bpoints table}. Dimensionless parameters are those given in table \ref{ethanol-water dimensionless base parameters table}.}
	\label{simulation loglog spreading graph}
\end{figure}

Taking a closer look at the influence of initial ethanol concentration on the spreading rate, figure \ref{simulation loglog spreading graph} plots radius growth versus time on a logarithmic scale for the data shown in figure \ref{ethwater Xa t graph}. As we know, the spreading behaviour of wetting droplets tends to obey a power law growth of radius in time, $r \propto t^n$, where $n$ is the spreading exponent. Therefore, the gradient of the radii plotted in figure \ref{simulation loglog spreading graph} will give the spreading exponents of for each $\chi_{A0,i}$. Note that similar values of $n$ can be found for the retraction rate. We can see from figure \ref{ethwater Xa t graph} that as we increase initial ethanol concentration, the line growth gradients and hence spreading exponents approach values of unity, moving into the realms of superspreading liquids such as droplets laden with trisiloxane surfactants \citep{Rafai2002, Karapetsas2011, Theodorakis2015}. 

Table \ref{modelling spreading coeffs bpoints table} gives the precise values for the linear fit. As with the experimental values (see table \ref{spreading const rmax T = 70C table}), $n_1$ gives the first spreading coefficient until the first breakpoint in time, $b_1$, where the gradient shifts to $n_2$ until time $b_2$ and so on until dry-out. We see that for pure water, $\chi_{A0,i} = 0.00$, there is an initial contact line adjustment with rapid spreading at early times where $n_1 = 0.6$. This value is close to the reported value by \citep{Winkels2012} $n = 0.55$ and within the range of the experimental error. The spreading exponent soon slows and settles at $n_2 = 0.11$, close to Tanner's law as expected for pure liquids \citep{Cazabat1986,Chen1989,Chen1987}. After time $b_3 = 0.78$, an exponent close to zero, $n_3 = 0.02$, shows a region where forces at the contact line are largely balanced and is effectively stationary before evaporation taking over and the droplet receding at increasing rates from $n_4$ to $n_8$. For the majority of the retraction time, $t =$ \SIrange{20.83}{34.24}{}, is conducted at exponent $n_6 = -0.50$. This is similar to retraction rates reported by \citet{Cachile2002,Cachile2002a} as well as \citet{Poulard2003}. The increasing retraction rate is explained by the shrinkage in droplet height from mass loss as it evaporates. As previously discussed, the reduced droplet thickness gives rise to greater evaporation rates since the droplet is heated more by the substrate.

\begin{table}
\begin{center}
\def~{\hphantom{0}}
\setlength{\tabcolsep}{8pt}
\begin{tabular}{lS[table-format=2.2]S[table-format=2.2]S[table-format=2.2]S[table-format=2.2]S[table-format=2.2]}
\multicolumn{1}{c}{} &\multicolumn{5}{c}{$\chi_{A0,i}$}\\[3pt]
	  &0               &0.10             &0.25           &0.50          &0.75\\[6pt]
$n_1$ &0.6             &0.45	         &0.5	         &1.12          &1.47   \\
$b_1$ &0.11            &0.26	         &0.54	         &0.15          &0.12   \\[4pt]
$n_2$ &0.11            &0.15	         &0.19	         &0.67          &0.89   \\
$b_2$ &0.78            &1.03	         &1.90	         &0.51          &0.35   \\[4pt]
$n_3$ &0.02            &0.05	         &-0.02	         &0.36          &0.51   \\
$b_3$ &2.54            &2.18	         &3.66	         &1.21          &0.80   \\[4pt]
$n_4$ &-0.05           &-0.12	         &-0.23	         &0.16          &0.27   \\
$b_4$ &8.75            &13.93	         &5.68           &2.31          &1.64   \\[4pt]
$n_5$ &-0.17           &-0.24	         &-0.39	         &0.00          &0.11   \\
$b_5$ &20.83           &21.86	         &8.12	         &3.44          &2.72   \\[4pt]
$n_6$ &-0.50           &-0.46	         &-0.65	         &-0.15         &-0.07  \\
$b_6$ &34.24           &30.62	         &10.12	         &4.61          &3.85   \\[4pt]
$n_7$ &-1.39           &-0.93	         &-0.30	         &-0.31         &-0.30  \\
$b_7$ &43.88           &38.99	         &26.48	         &6.11          &5.11   \\[4pt]
$n_8$ &-4.18           &-2.14	         &-1.22	         &-0.45         &-0.60  \\
\end{tabular}
\caption{Predicted spreading exponents, $n$ and corresponding breakpoints in time, $b$ for increasing initial concentrations of ethanol, $\chi_{A0,i}$.} \label{modelling spreading coeffs bpoints table}
\end{center}
\end{table}
To reveal more information about the flow field, we decompose the averaged velocity at the interface, $u$, into three distinct components,
\begin{equation}
u = u_{tg} + u_{cg} + u_{ca}
\end{equation}
These are the three mechanisms that can drive movement and spreading of the contact line: $u_{tg}$ is the thermocapillary velocity, where surface tension gradients arising from temperature variations drive the fluid motion; $u_{cg}$ is the solutocapillary velocity, where flow is driven by a surface tension gradient sustained by an uneven mixture concentration; and, $u_{ca}$ is the capillary velocity, sustained by the capillary pressure over the interface. By decomposing the bulk velocity into these three contributions, we can gain insight into the driving forces governing the spreading behaviour. It can be shown that for the limiting case of $Re=0$, the decomposed velocities at the interface are expressed as,
\begin{equation}
u_{ca} = - \frac{h^2} {2 \mu} \frac{\partial p}{\partial r}
\end{equation}
\begin{equation}
u_{cg} = \bigg[\frac{\partial \chi_{A0}}{\partial r} - \sigma_R \frac{\partial \chi_{A0}}{\partial r} - Ma T_s \frac{\partial \chi_{A0}}{\partial r} ( 1 - \gamma_R ) \bigg]\frac{h}{\mu Ma}
\end{equation}
\begin{equation}
u_{tg} = \bigg[-\frac{\partial T_s}{\partial r} \chi_{A0} -  \frac{\partial T_s}{\partial r} \gamma_R(1-\chi_{A0})\bigg] \frac{h}{\mu}  
\end{equation}
The roles of these components will be discussed in detail for various cases in the following sections.
\subsubsection{Low initial ethanol concentration}
\begin{figure}
	\centering
	\includegraphics[width=0.5\textwidth]{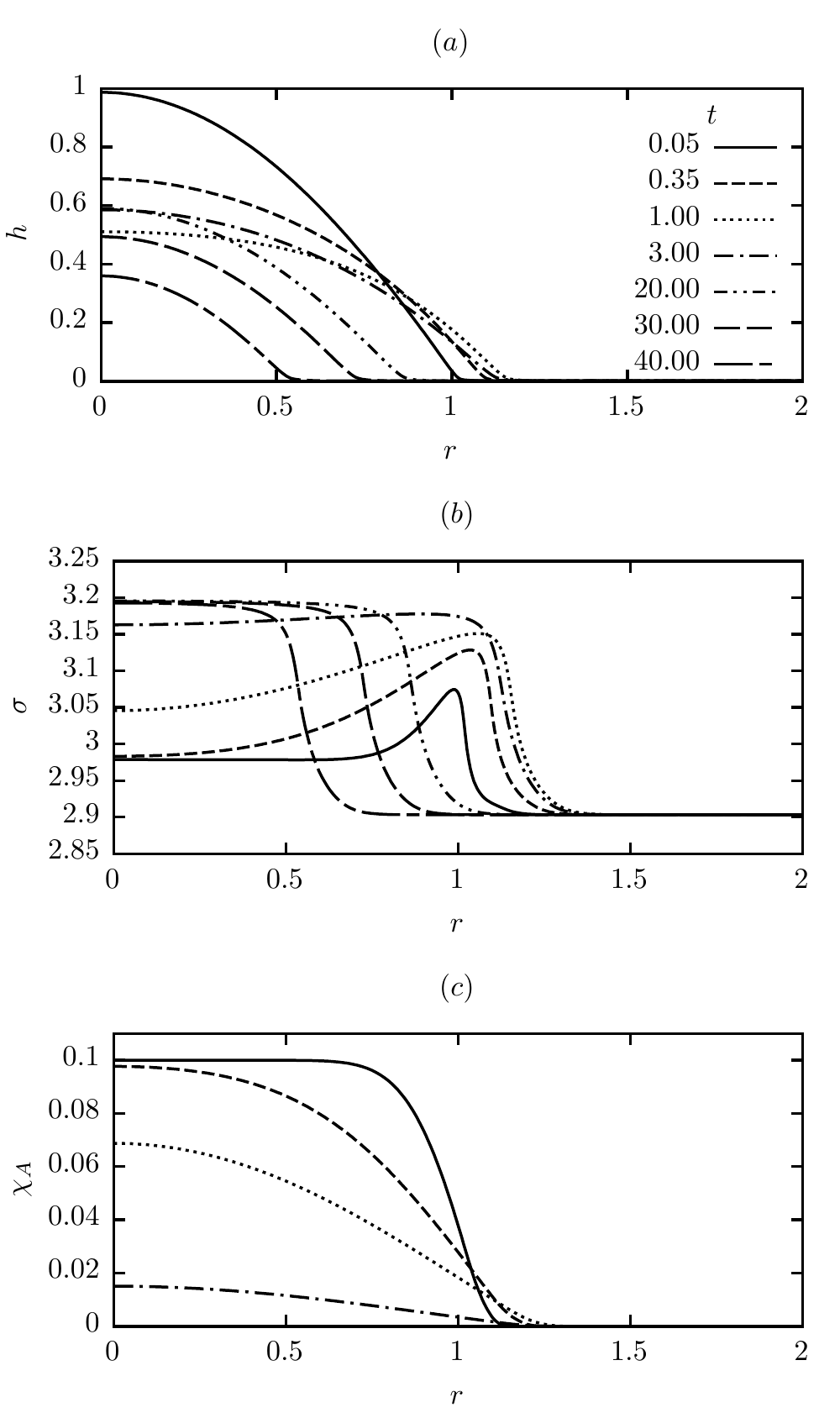}
	\caption{Snapshots of (a) interface profile, (b) surface tension, and (c) concentration of component $A$ for an ethanol-water droplet with $\chi_{A0,i} = 0.10$. Dimensionless parameters are those given in table \ref{ethanol-water dimensionless base parameters table}.}
	\label{Xa=010 h profile snapshots}
\end{figure}
\begin{figure}
	\centering
	\includegraphics[width=0.5\textwidth]{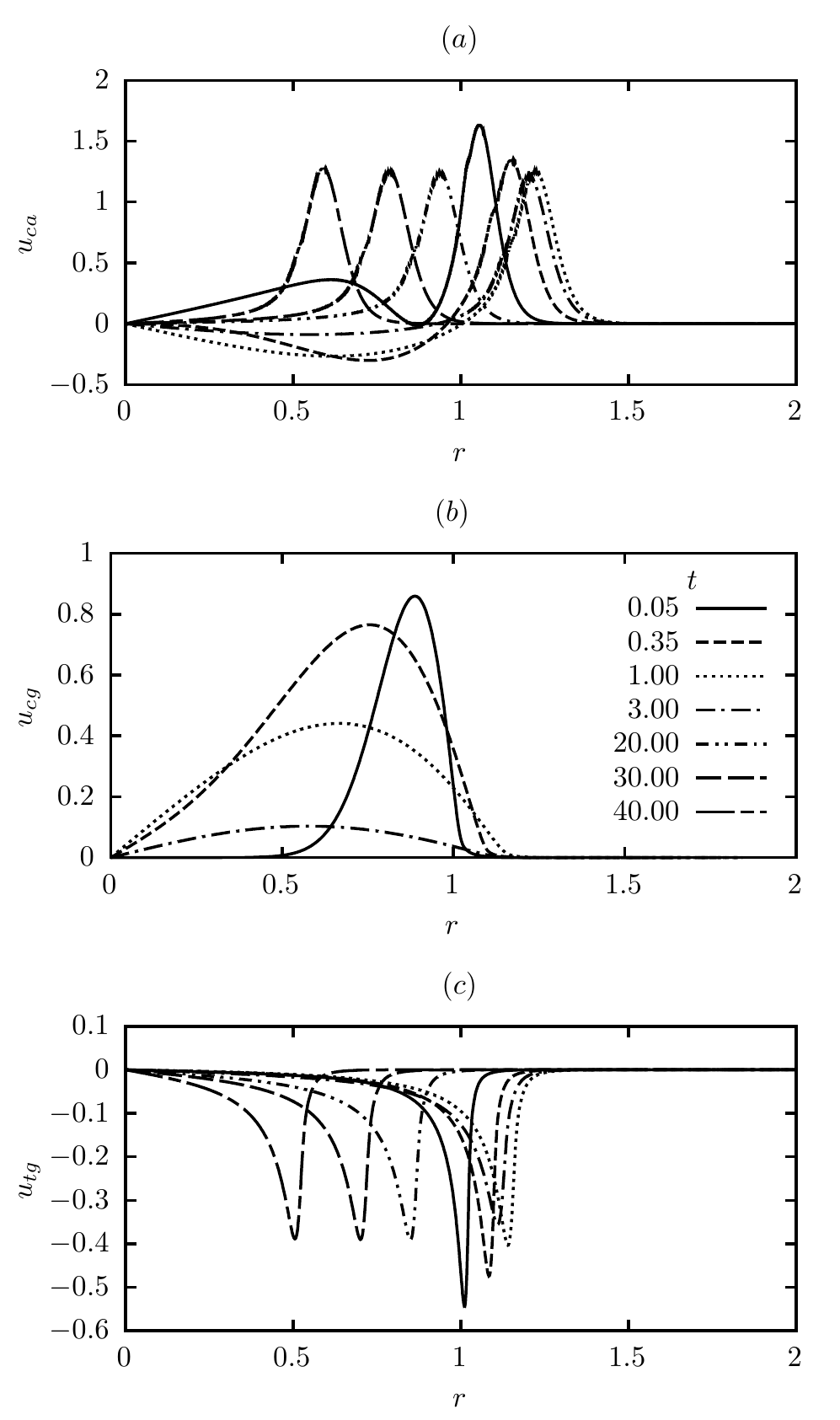}
	\caption{Snapshots of decomposed surface velocities for an ethanol-water droplet with $\chi_{A0,i} = 0.10$ over its lifetime. (a) capillary velocity, (b) solutocapillary velocity, (c) thermocapillary velocity. Dimensionless parameters are those given in table \ref{ethanol-water dimensionless base parameters table}.}
	\label{Xa=010 Udecomp snapshots}
\end{figure}
Figure \ref{Xa=010 h profile snapshots} shows the evolution of interface position, surface tension and ethanol mass fraction along $r$ for an ethanol-water droplet with $\chi_{A0,i} = 0.10$. The interface profile, figure \ref{Xa=010 h profile snapshots}(a), indicates that the droplet spreads significantly between $t = 0.05$ and $t = 0.35$ with a significant droplet in apex height of \num{0.3}. From table \ref{modelling spreading coeffs bpoints table}, we can see that $n_2$ rises to \num{0.15} with the increased spreading rate lasting for longer times until $b_2 = 1.03$. It must be noted that for $\chi_{A0,i} = 0.25$, $n_2 = 0.19$ until $b_2 = 1.90$. This trend was also seen by \citet{Guena2007b} when increasing concentration of the more volatile alkane. Figure \ref{Xa=010 h profile snapshots}(b) reveals that the surface tension gradient between the apex and contact line increases during this period with figure \ref{Xa=010 h profile snapshots}(c) showing increased depletion of ethanol closer to the contact line. Spreading continues until $t = 1$ and by $t = 3$, the droplet begins to recede as thermal Marangoni effects start to dominate. The apex height increases from $t = 1$ as thermal Marangoni stress pulls liquid towards the centre. Inspection of figure \ref{Xa=010 h profile snapshots}(c) shows that ethanol is still present within the droplet in small amounts ($\chi_{A0} < 0.02$). If we compare the breakpoint time $b_2$ signifying the end of the spreading regime with Fig \ref{ethwater Xa t graph}(d) showing apex ethanol mass fraction, we see that ethanol is not totally depleted within the droplet until $t = 10$ in both cases. This suggests that a residual amount of ethanol remains in the droplet well into the recession regime. By the next snapshot, at $t = 20$, ethanol is totally depleted in the droplet and evaporation now proceeds relatively slowly with the interface retaining a spherical cap shape. We can see in figure \ref{Xa=010 h profile snapshots}(b) that surface tension at later times is always higher at the apex, however, the magnitude of the surface tension gradient is significantly smaller than the reverse gradient present at early times due to concentration effects. 

We now examine the decomposed interface velocities of these time snapshots in figure \ref{Xa=010 Udecomp snapshots}. A positive value indicates velocity directed towards the contact line while a negative value shows velocity directed towards the centre. Capillary velocity, $u_{ca}$, resulting from interface curvature is predictably large and positive at the contact line as the droplet profile transitions into the precursor layer while becoming negative towards the centre due to reverse curvature. Figure \ref{Xa=010 Udecomp snapshots}(a) shows the movement of $u_{ca}$ over time with the spreading and recession of the contact line. The solutocapillary velocity, $u_{cg}$, in figure \ref{Xa=010 Udecomp snapshots}(b) displays a clear trend. It is positive at all times, driving liquid towards the contact line and decays over time; $u_{cg}$ is largest at the earliest time of $t =0.05$ when the concentration gradient between the apex and contact line is also at its greatest. The strength of the outward solutocapillary velocity gradually decreases as $\chi_{A0}$ evaporates until beyond $t = 3.00$ where it decays completely---coinciding with total depletion of $\chi_{A0}$. Figure \ref{Xa=010 Udecomp snapshots}(c) tracks the development of the theromocapillary velocity, $u_{tg}$, which is negative at all times. Again, this is in line with the work of \citet{Ajaev2005} and \citet{Ehrhard1991} by demonstrating that thermocapillary force is partly responsible (aside from evaporative cooling and heat transfer from the substrate) for forcing the fluid inwards towards the droplet centre. The largest magnitude of $u_{tg}$ is always located at the contact line, becoming more negative the thinner the film becomes, corresponding to a warmer region.

\begin{figure}
	\centering
	\includegraphics[width=0.55\textwidth]{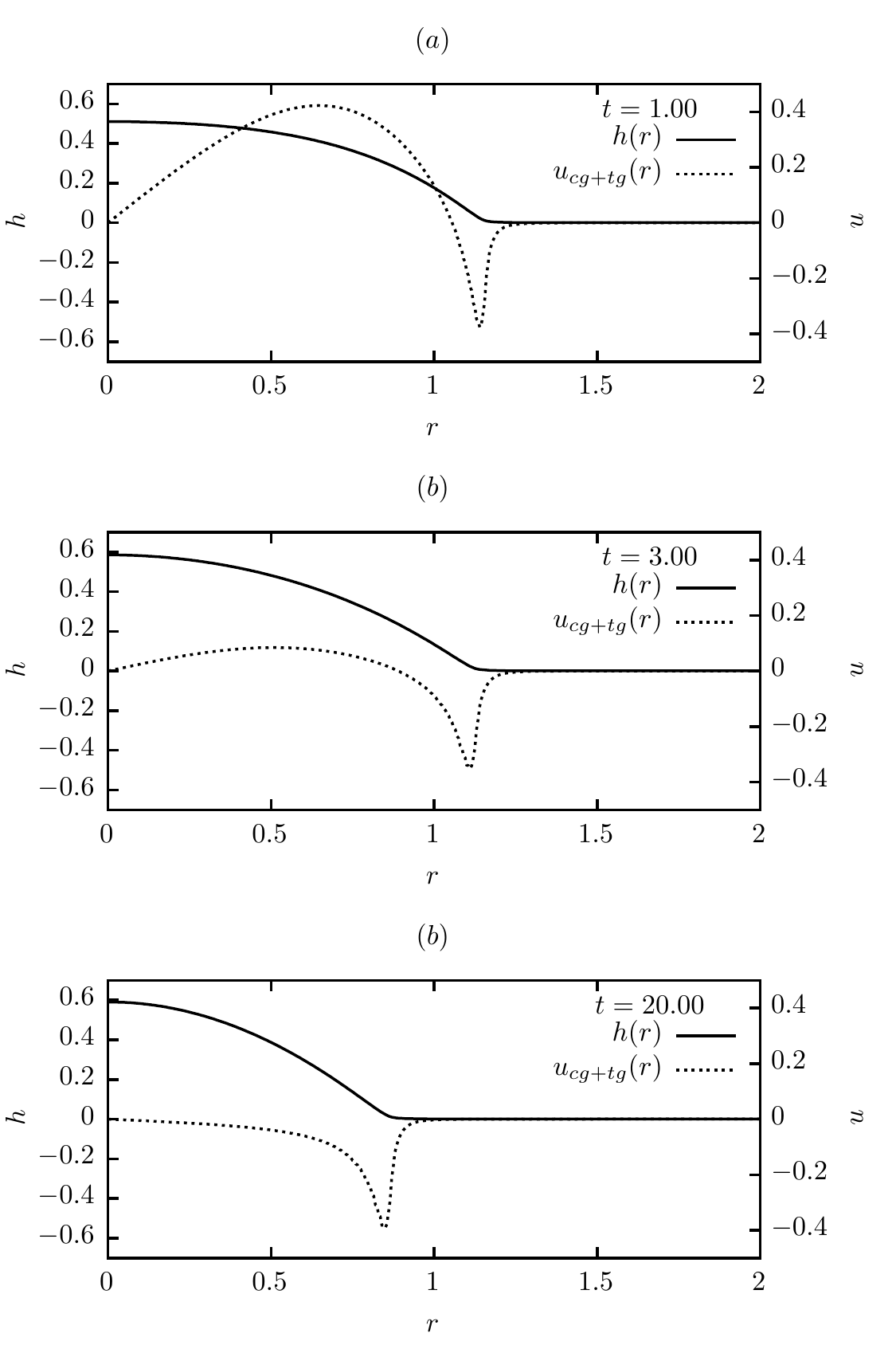}
	\caption{Interface profile and corresponding combined Marangoni velocity (solutal and thermal) for an ethanol-water droplet with $\chi_{A0,i}=0.10$. Other dimensionless parameters are those given in table \ref{ethanol-water dimensionless base parameters table}. (a)$t = 1.00$, (b) $t = 3.00$, (c) $t = 20.00$}
	\label{Xa=010 Ucg+Utg snapshots}
\end{figure}
Examining further the balance between thermal and solutal Marangoni stresses, we turn our attention to figure \ref{Xa=010 Ucg+Utg snapshots} which illustrates the combined Marangoni velocity profiles at times $t = 1$, $t = 3$, and $t = 20$, along with the interface profile. The droplet radius is largest at $t = 1$ before beginning to recede at $t = 3$. Figure \ref{Xa=010 Ucg+Utg snapshots}(a) shows a net negative (inward) Marangoni velocity in the vicinity of the contact line with a net positive (outward) velocity in the droplet interior. As time proceeds, $u_{cg}$ diminishes in strength and so this action combined with the constant inward flow of $u_{tg}$ halts the movement of the contact line. By $t = 3$, $\chi_{A0}$ is sufficiently depleted that there is only a weak outward combined Marangoni velocity in the bulk droplet with the overwhelming velocity directed inwards from the contact line. By $t = 20$, the combined Marangoni velocity throughout the whole droplet profile is negative and directed inwards with the absence of any solutal effects.
\subsubsection{High initial ethanol concentration}
\begin{figure}
	\centering
	\includegraphics[width=0.6\textwidth]{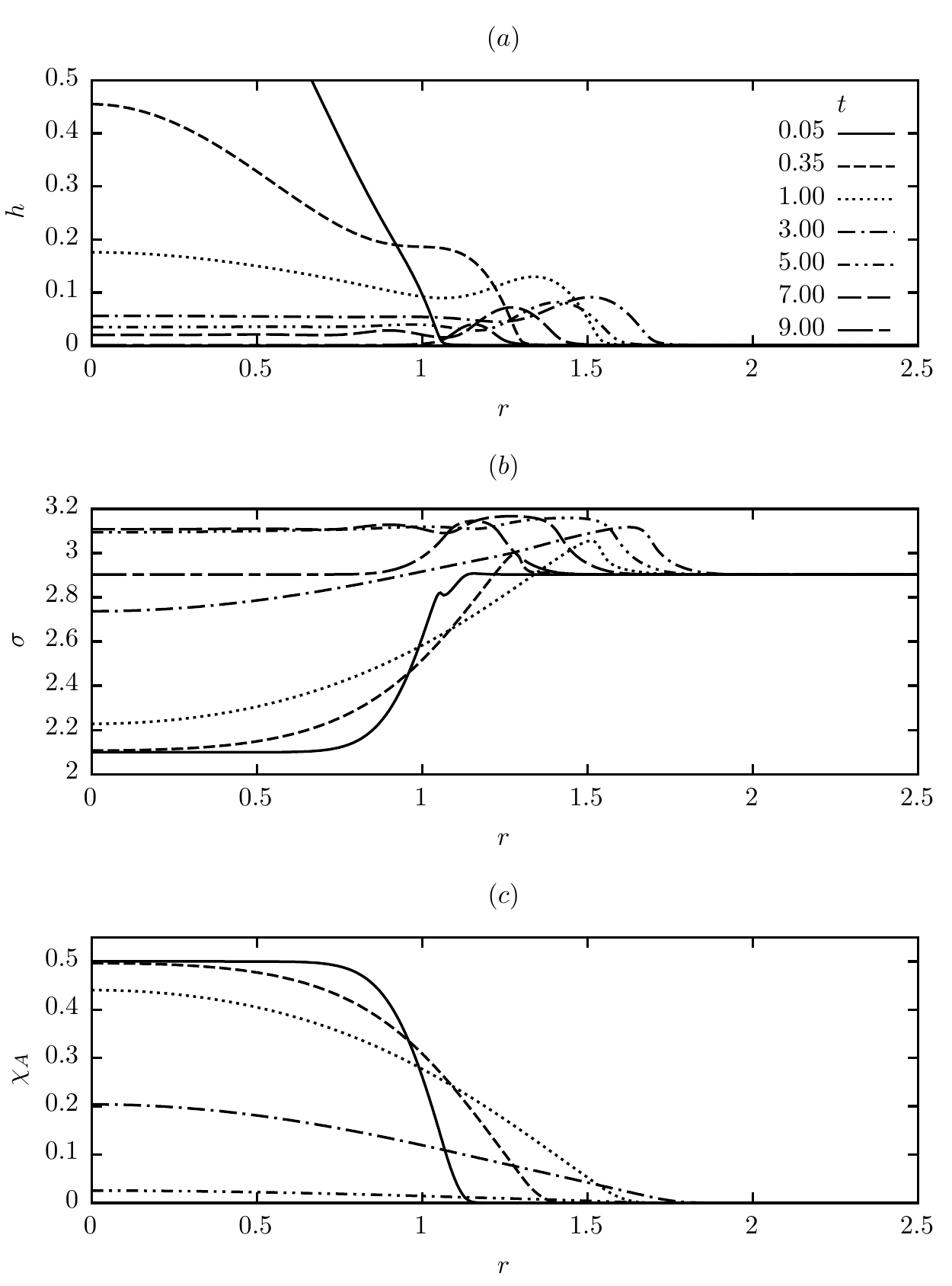}
	\caption{Snapshots of (a) interface profile, (b) surface tension, and (c) concentration of component $A$ along the interface for an ethanol-water droplet with $\chi_{A0,i} = 0.50$. Dimensionless parameters are those given in table \ref{ethanol-water dimensionless base parameters table}.}
	\label{Xa=050 h profile snapshots}
\end{figure}
\begin{figure}
	\centering
	\includegraphics[width=0.6\textwidth]{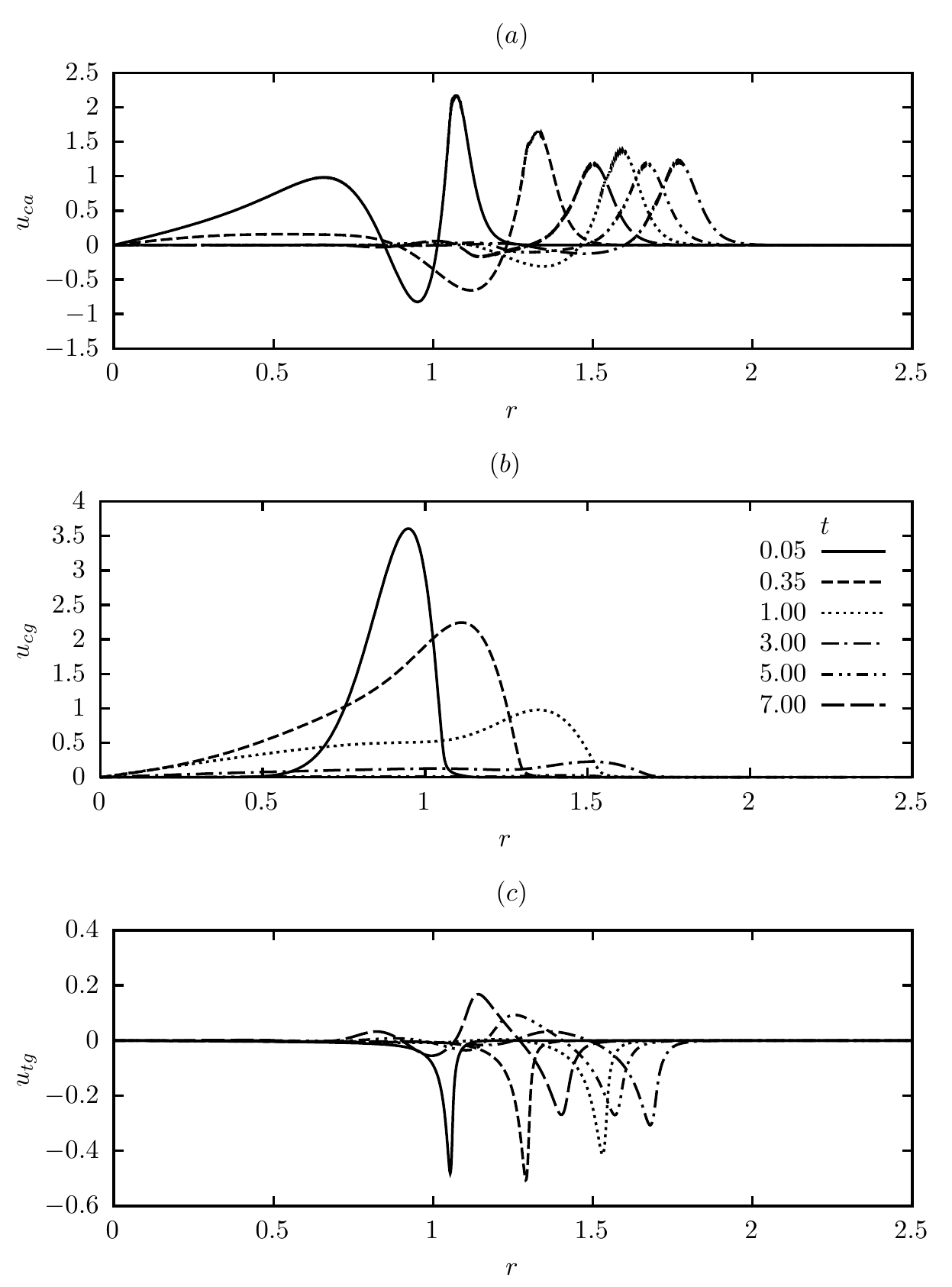}
	\caption{Snapshots of decomposed surface velocities for an ethanol-water droplet with $\chi_{A0,i} = 0.50$ over its lifetime. (a) capillary velocity, (b) solutocapillary velocity, (c) thermocapillary velocity. Dimensionless parameters are those given in table \ref{ethanol-water dimensionless base parameters table}.}
	\label{Xa=050 Udecomp snapshots}
\end{figure}
\begin{figure}
	\centering
	\includegraphics[width=0.65\textwidth]{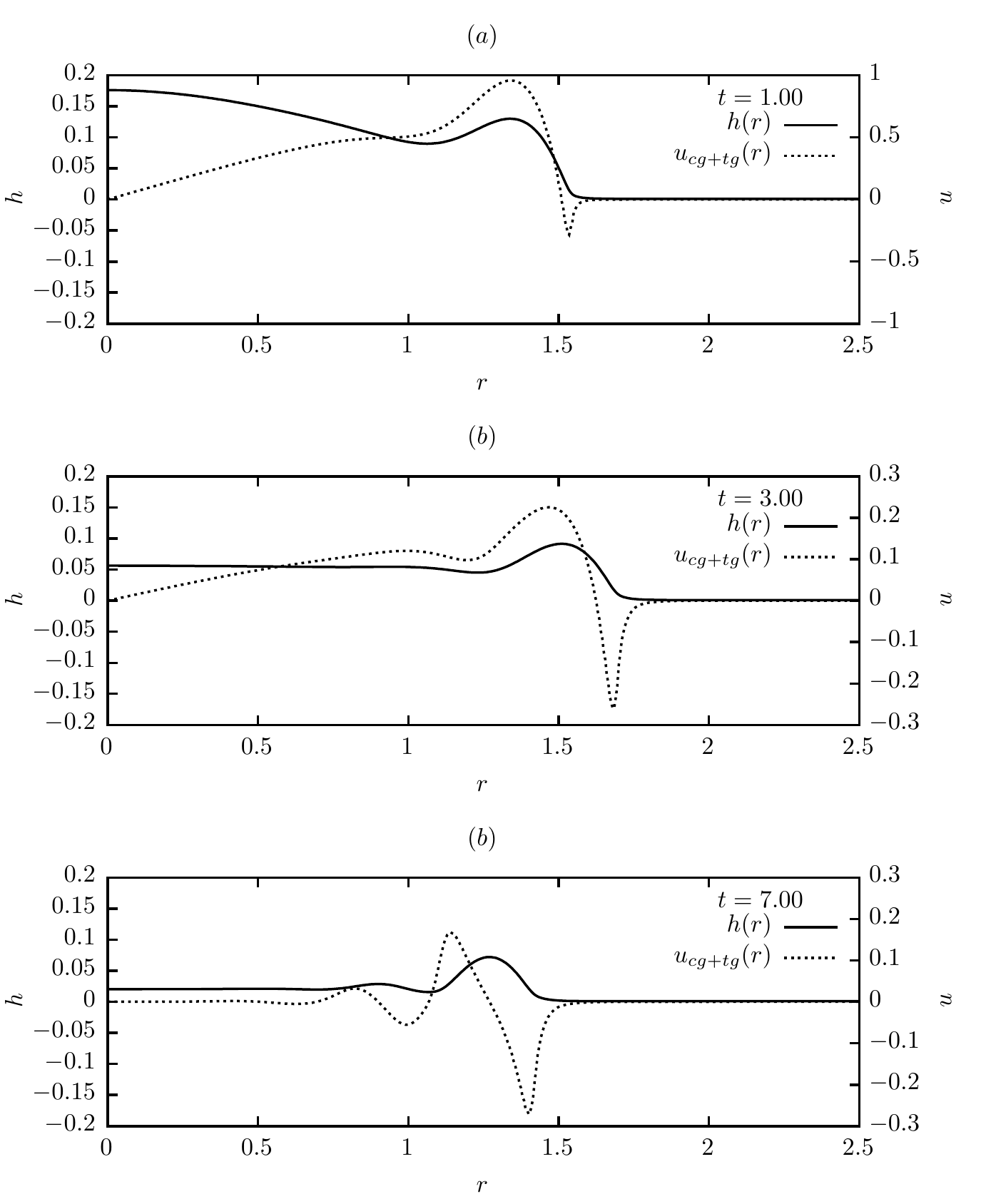}
	\caption{Interface profile and corresponding combined Marangoni velocity (solutal and thermal) for an ethanol-water droplet with $\chi_{A0,i}=0.50$. Other dimensionless parameters are those given in table \ref{ethanol-water dimensionless base parameters table}. (a)$t = 1.00$, (b) $t = 3.00$, (c) $t = 7.00$}
	\label{Xa=050 Ucg+Utg snapshots}
\end{figure}

When the initial ethanol concentration is increased to $\chi_{A0,i} = 0.50$, the evolution of the droplet profile becomes more complex. In figure \ref{Xa=050 h profile snapshots} we again examine the evolution of the interface position, surface tension and mass fraction of ethanol. With figures \ref{Xa=050 Udecomp snapshots} and \ref{Xa=050 Ucg+Utg snapshots} we explore the decomposed velocities in more detail. It is clear from figure \ref{Xa=050 h profile snapshots}(a) that evolution of the interface is different from $\chi_{A0,i} = 0.10$ in figure \ref{Xa=010 h profile snapshots}. From $t = 0.05$ to $t = 3.00$, the droplet spreads rapidly to a pancake shape with the formation of a ridge of liquid preceding the contact line. This is similar to the ridge formed in the spreading of trisiloxane-laden surfactant droplets \citep{Rafai2002, Karapetsas2011} and results from the rapid rate of spreading. Table \ref{modelling spreading coeffs bpoints table} shows that the first spreading exponent $n_2$ is now significantly higher at \num{0.67} with the rate progressively decreasing to $n_3 = 0.36$ and $n_4 = 0.16$ (closer to Tanner's law) before the contact line retracts. This is due to the decreasing concentration gradient between the contact line and apex as ethanol evaporates and solutal Marangoni stresses weaken. Figure \ref{Xa=050 h profile snapshots} reveals that before $t = 3$, surface tension is always largest towards the contact line, specifically at the apex of the ridge. The contact line can be seen retracting from $t = 5$ onwards while the flat plane in the droplet interior trapped by the ridge gradually decreases in height. Notice that at $t = 9$, the droplet centre has reached dry-out, however the ridge at the contact line still remains. Extrapolated in the azimuthal plane to three dimensions, film dry-out leaves a torus shaped ring of liquid. This is analogous to ring observed in the experiments conducted by \citet{Guena2007b} on droplets of alkane mixtures evaporating from isothermal substrates. Figure \ref{Xa=050 h profile snapshots}(c) confirms that all ethanol (component $A$) is depleted from the droplet by $t = 7.00$ and so it can be concluded that the ridge consists entirely of water (component $B$). Similar behaviour is also seen at $\chi_{A0,i} = 0.75$ (not shown), however with a greater initial rate of $n_2 = 0.89$ and the emergence of three further distinct linear spreading regimes: $n_3 = 0.51$, $n_4 = 0.27$, and $n_5 = 0.11$. Overall retraction exponents decrease with increasing $\chi_{A0,i}$. As will be explained later, this is owing to the increased solutal Marangoni outward force acting against inward thermal Marangoni stresses.

In figure \ref{Xa=050 Udecomp snapshots}(a) we see that $u_{ca}$ is larger than the $\chi_{A0,i} = 0.10$ case at early times. $u_{ca}$ is largest at the contact line at all times, even during ridge formation. A similar trend is displayed in solutocapillary velocity as before, the key difference being that the magnitude of $u_{cg}$ is around four times larger when $\chi_{A0,i} = 0.50$ over $\chi_{A0,i} = 0.10$. This is expected due to the higher concentration gradient between the apex and contact line. It also appears from figure \ref{Xa=050 Udecomp snapshots}(b) that outward flow from $u_{cg}$ is negligible at $t = 3.00$ and this is the time at which retraction begins. The thermocapillary velocities in figure \ref{Xa=050 Udecomp snapshots} show an altogether more interesting trend. Before ridge formation, $u_{tg}$ is of the same direction and magnitude as the $\chi_{A0,i} = 0.10$ case---around \num{0.5} directed inwards toward the droplet centre. However, as the droplet flattens and the ridge forms, a positive $u_{tg}$ begins to emerge on the LHS of the ridge. This velocity pushes fluid from the bulk droplet outwards toward the ridge while there is simultaneously a negative $u_{tg}$ on the RHS of the ridge pushing fluid inward. Physically, this means that liquid from both sides is flowing towards the ridge, sustaining its formation. As liquid flows from the thin plane on the LHS to feed the ridge, the removal of liquid from the thin layer causes a dimple in the interface profile to form adjacent to the ridge. This can be seen by examining $h$ in figure \ref{Xa=050 h profile snapshots}(a) from $t = 5.00$ to $t = 7.00$ to $t = 9.00$ where the ridge is shown steadily receding while the interior dries out. The reduced thickness of the interface in this region causes the liquid to be heated to a greater temperature and hence produces a larger surface tension gradient between the bottom of the dimple and the apex of the ridge. This then results in a stronger thermocapillary velocity from the dimple to the ridge which can be seen clearly in figure \ref{Xa=050 Udecomp snapshots}(c). Therefore, it appears that the initial ridge is formed due to solutocapillarity inducing very rapid spreading of the contact line. Once formed, the ridge is sustained by thermocapillarity providing a steady flow of fluid to the apex. 

Finally, let us consider the combined actions of the solutal and thermal Marangoni velocities at key points in the $\chi_{A0,i} = 0.50$ droplet lifetime. Figure \ref{Xa=050 Ucg+Utg snapshots}(a) shows the interface profile and combined Marangoni velocity at $t = 1$ while the droplet is still firmly in the spreading regime. Figure \ref{Xa=050 Ucg+Utg snapshots}(b) considers $t = 3.00$ when maximum radius is reached and (c) shows the droplet well into the recession regime at $t = 7.00$, with the liquid film on the LHS of the ridge still present but close to dry-out. At $t = 1$, velocity is overwhelmingly directed towards the contact line with a small inward velocity at the contact line itself where liquid is warmest. Inward velocity at the contact line grows by $t = 3$ while outward velocity declines as ethanol evaporates. By $t = 7.00$, there is a clear inward Marangoni velocity from the RHS of the ridge as the droplet contact line recedes. The dimple in the interface profile on the LHS of the ridge is also visible. At the minimum point of the dimple, there is a positive and negative velocity on either side (the RHS and LHS respectively). This means that fluid from the dimple is driven both outwards towards the ridge at the contact line and inward towards the centre. The mechanism sustains ridge formation even after spreading has finished and only water remains in the droplet. The simultaneously decreasing dimple depth increases the strength of the Marangoni flow while intimately leading to dry-out in the interior before the contact line ridge completely evaporates.
\section{Parametric analysis}
As reported by \citet{Guena2007b}, the spreading of small binary mixture sessile droplets is a complex process governed by a delicate interplay between evaporation, surface tension gradients, mass diffusion, hydrodynamic flow, and capillary forces. An explicit advantage of our model over experiments is the ability to alter specific dimensionless numbers while keeping other properties constant, allowing us to assess the impact of each mechanism individually. We now briefly examine the effect changing the magnitude of $E$, $K$, $Ma$, $\sigma_R$, $Pe$, and $Re$ on the solution on for $\chi_{A0,i} = 0.50$.
\subsection{Evaporation number}
\label{Evaporation number}
\begin{figure}
	\centering
	\includegraphics[width=0.75\textwidth]{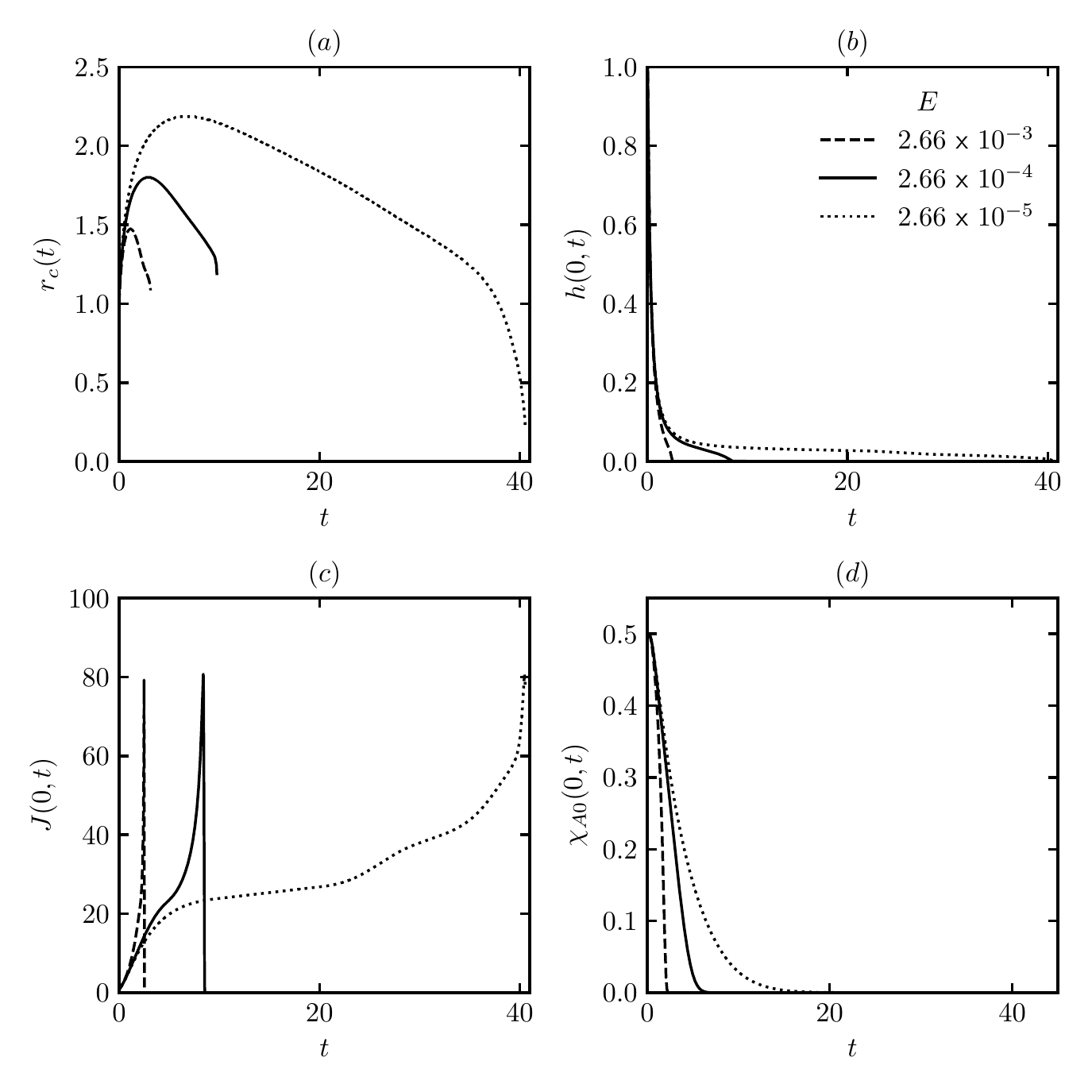}
	\caption{Profiles of (a) contact line position, (b) apex height, (c) apex mass flux, and (d) apex mass fraction throughout the lifetime of a $\chi_{A0,i} = 0.50$ droplet with varying Evaporation numbers, $E$. Unless otherwise stated, dimensionless parameters are those given in table \ref{ethanol-water dimensionless base parameters table}.}
	\label{compare E Xa = 050}
\end{figure}
Increasing evaporation number, $E$, increases the volatility of both components in the mixture and is hence analogous to increasing the substrate temperature in an experimental scenario. In figure \ref{compare E Xa = 050}, we examine the effect of increasing and then decreasing $E$ by one order of magnitude over the base case value of $E =$ \num{2.66e-4} given in table \ref{ethanol-water dimensionless base parameters table}. Increasing $E$ to \num{2.66e-3} simultaneously reduces spreading extent and droplet lifetime as evaporation rate of both liquids becomes larger. Decreasing $E$ to \num{2.66e-5} (analogous to lowering the substrate temperature) has the opposite effect. With evaporation now weaker, the droplet spreads to a larger maximum radius where it remains stationary for a period before retraction. These trends are similarly reflected in the profiles of evaporative flux and ethanol mass fraction as the droplet apex shown in figures \ref{compare E Xa = 050}(c) and (d) respectively. We see a similar trend here as we do in our experimental findings when substrate temperature is varied---see section \ref{Variation in temperature}.
\subsection{Knudsen number}
\label{Knudsen number}
\begin{figure}
	\centering
	\includegraphics[width=0.75\textwidth]{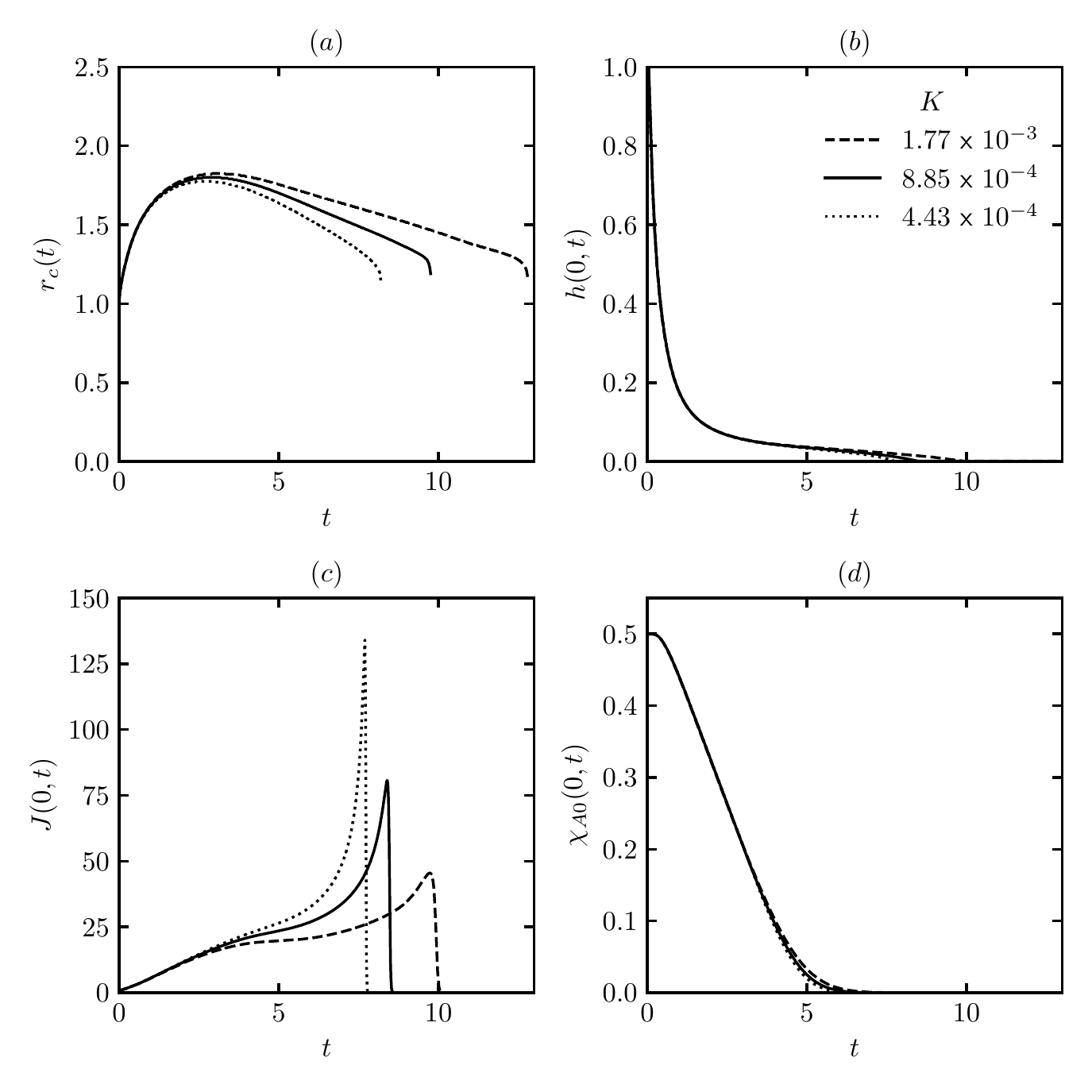}
	\caption{Profiles of (a) contact line position, (b) apex height, (c) apex mass flux, and (d) apex mass fraction throughout the lifetime of a $\chi_{A0,i} = 0.50$ droplet with varying Knudsen numbers, $K$. Unless otherwise stated, dimensionless parameters are those given in table \ref{ethanol-water dimensionless base parameters table}.}
	\label{compare K Xa = 050}
\end{figure}
The Knudsen number, $K$, measures the degree of nonequilibrium at the evaporating interface. Increasing $K$ decreases the heat transfer rate across the interface, causing the mixture to evaporate more slowly, hence having the the opposite effect to increasing $E$. This is shown in figure \ref{compare K Xa = 050} where we double and half the base case value of $K=$ \num{8.55e-4} from table \ref{ethanol-water dimensionless base parameters table}. Figure \ref{compare K Xa = 050}(c) clearly illustrates that as $K$ is increased, the total evaporative flux at the drop apex decreases, slowing contact line retraction and extending the lifetime of the droplet. 
\subsection{Marangoni number}
\begin{figure}
	\centering
	\includegraphics[width=0.75\textwidth]{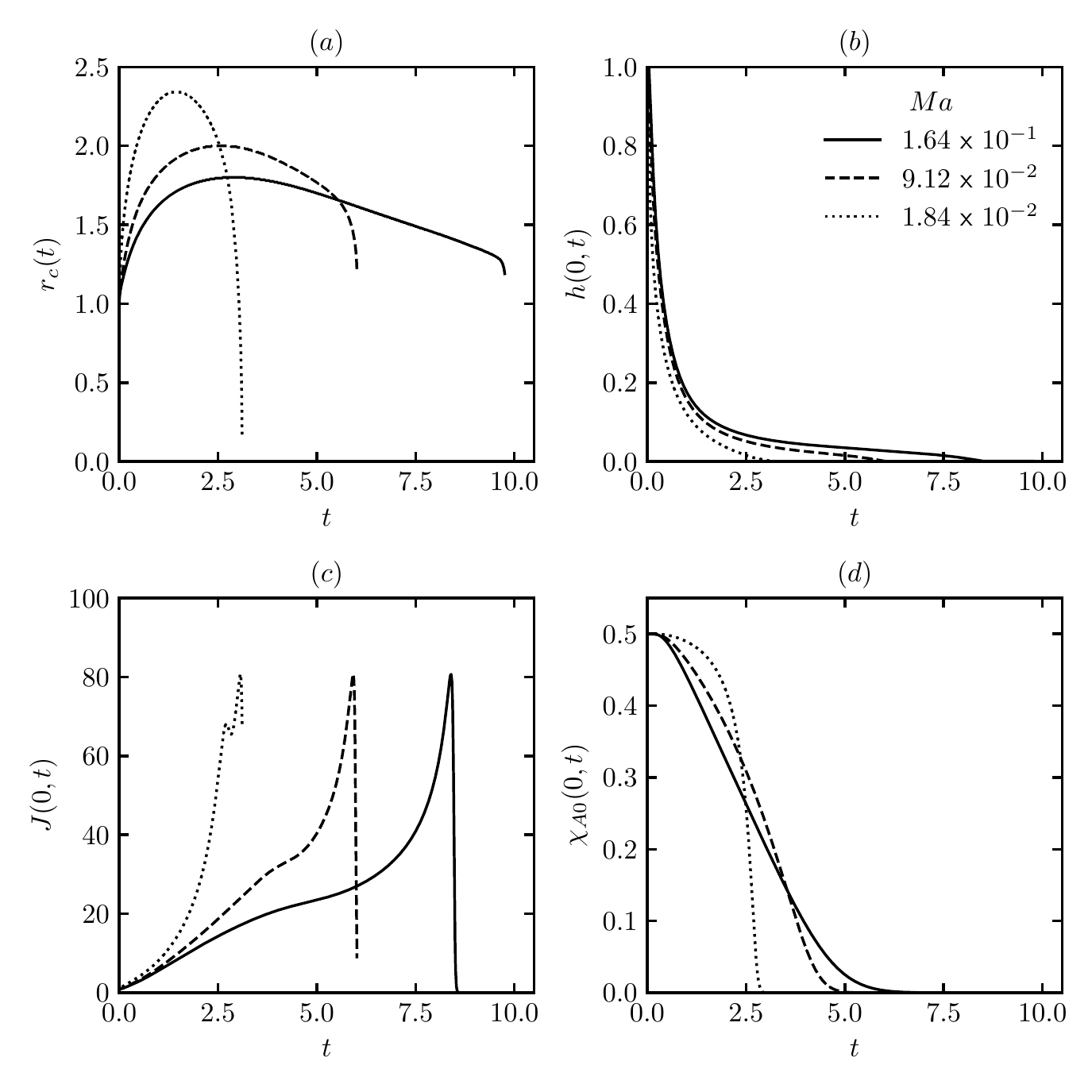}
	\caption{Profiles of (a) contact line position, (b) apex height, (c) apex mass flux, and (d) apex mass fraction throughout the lifetime of a $\chi_{A0,i} = 0.50$ droplet with varying Marangoni numbers, $Ma$. Unless otherwise stated, dimensionless parameters are those given in table \ref{ethanol-water dimensionless base parameters table}.}
		\label{compare Ma Xa = 050}
\end{figure}
The Marangoni number controls the strength of thermal Marangoni forces and hence the thermocapillary velocity, $u_{tg}$. We progressively decrease the base case value of $Ma =$ \num{1.64e-1} to \num{9.12e-2} and then \num{1.84e-2}, gradually weakening the thermal Marangoni stress. We see from figure \ref{compare Ma Xa = 050} that reducing $Ma$ increases the spreading rate and maximum droplet radius. This can be explained by the reduction of inward velocity $u_{tg}$ which provides opposition to spreading. Droplets that spread further are thinner films leading to greater evaporative flux---see figures \ref{compare Ma Xa = 050}(b) and (c). This ultimately leads to a shorter droplet lifetime at lower $Ma$. 
\subsection{Surface tension ratio}
\begin{figure}
	\centering
	\includegraphics[width=0.75\textwidth]{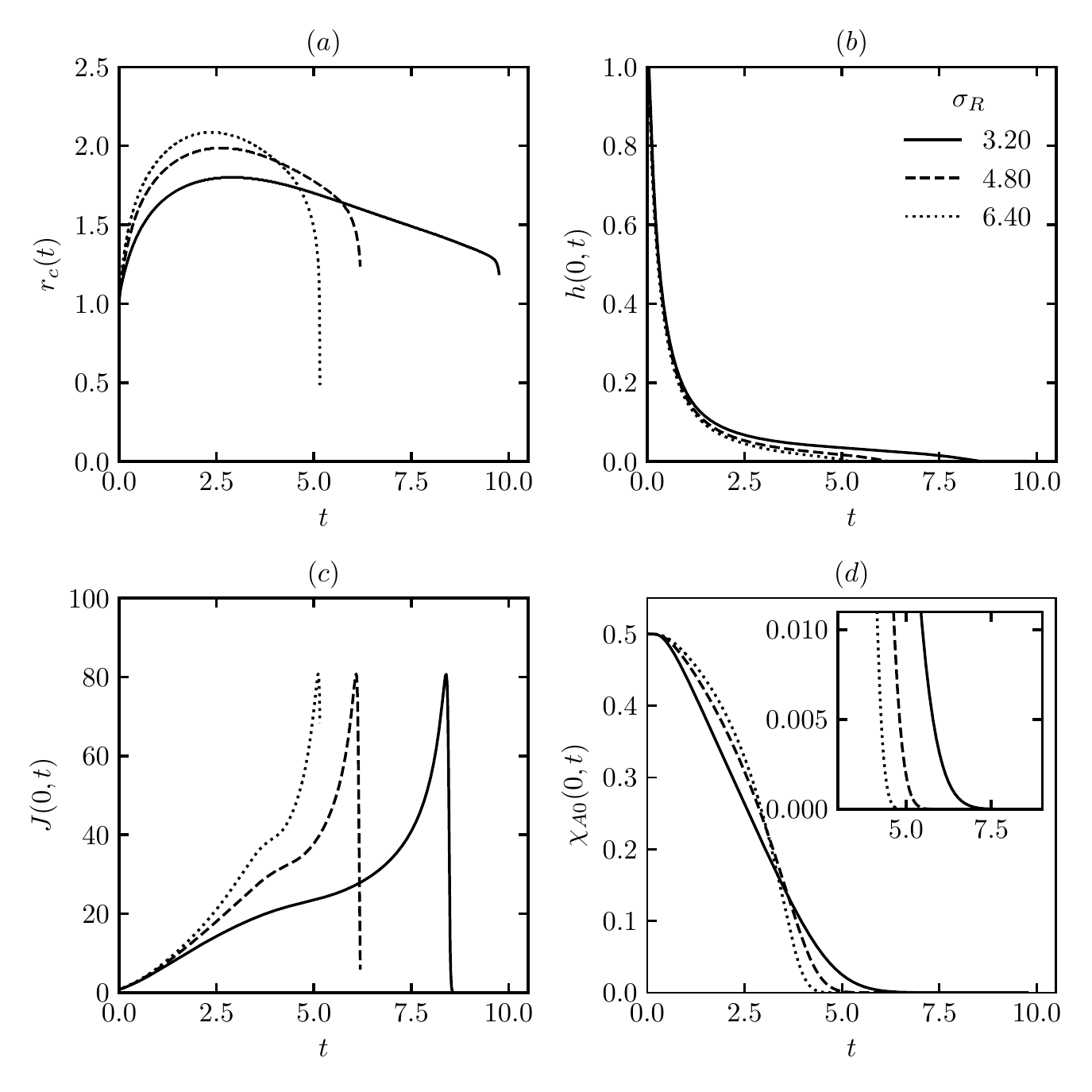}
	\caption{Profiles of (a) contact line position, (b) apex height, (c) apex mass flux, and (d) apex mass fraction throughout the lifetime of a $\chi_{A0,i} = 0.50$ droplet with varying surface tension ratio, $\sigma_R$. Unless otherwise stated, dimensionless parameters are those given in table \ref{ethanol-water dimensionless base parameters table}.}
	\label{compare SR Xa = 050}
\end{figure}
By increasing the surface tension ratio, $\sigma_R$, we can strengthen solutal Marangoni forces in the droplet. Larger $\sigma_R$ means the surface tension of the LVC is increased relative to the MVC. When $\chi_{A0,i} = 0.50$, as in figure \ref{compare SR Xa = 050}, the concentration induced surface tension gradient becomes larger as $\sigma_R$ increases. The larger surface tension gradient will amplify the outward solutocapillary velocity, $u_{cg}$, with liquid being more strongly drawn toward the contact line. Similar to cases with lowered Marangoni numbers, the increased spreading results in a thinner droplet subject to higher evaporative fluxes, hence resulting in shorter lifetimes.
\subsection{P\'{e}clet number}
\begin{figure}
	\centering
	\includegraphics[width=0.75\textwidth]{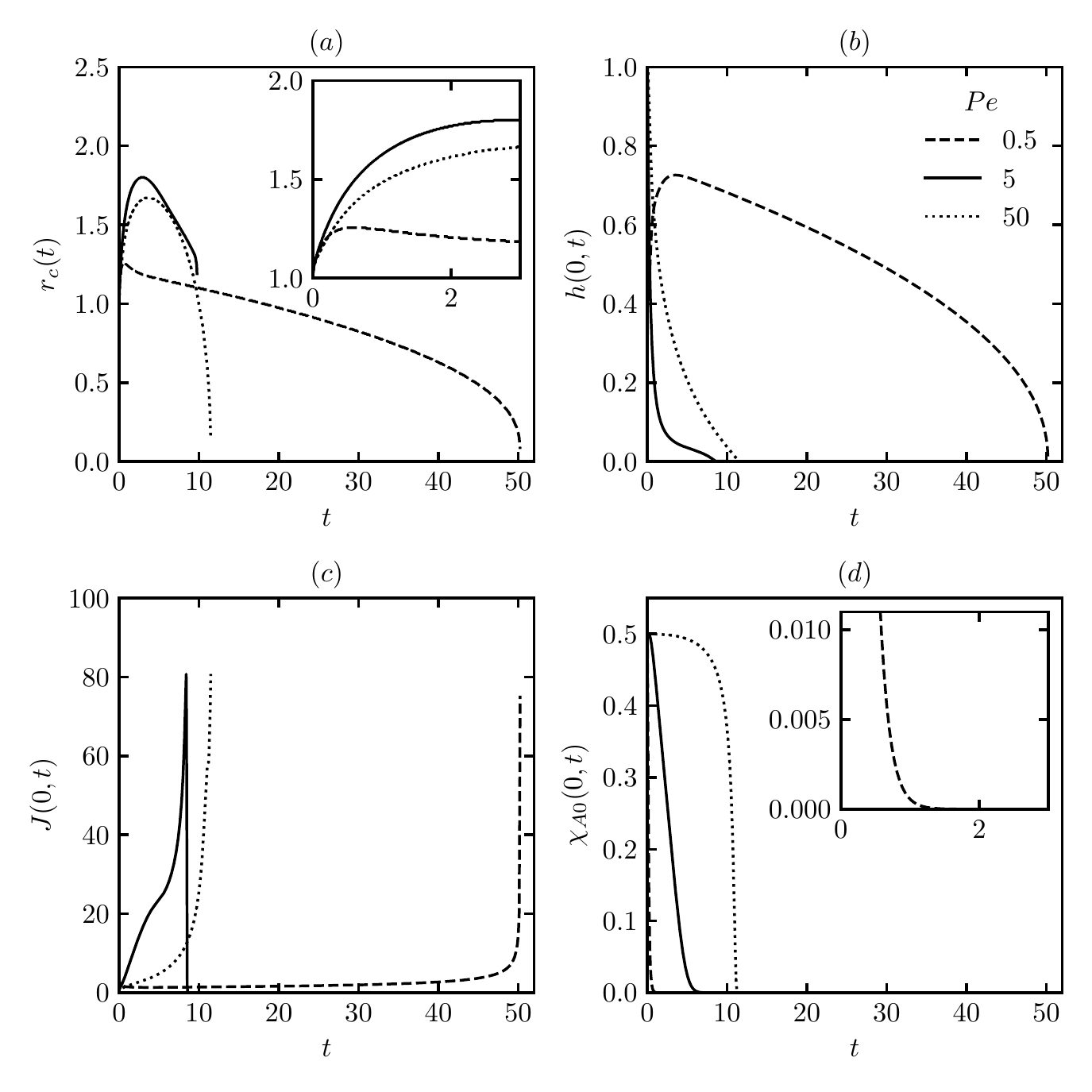}
	\caption{Profiles of (a) contact line position, (b) apex height, (c) apex mass flux, and (d) apex mass fraction throughout the lifetime of a $\chi_{A0,i} = 0.50$ droplet with varying P\'{e}let numbers $Pe$. Unless otherwise stated, dimensionless parameters are those given in table \ref{ethanol-water dimensionless base parameters table}.}
	\label{compare Pe Xa = 050}
\end{figure}
The mass diffusion is controlled by the P\'{e}clet number, with smaller values signifying more rapid diffusion of the MVC, ethanol in our case. By default, the base value in Table \ref{ethanol-water dimensionless base parameters table} is set to $Pe = 5$. In figure \ref{compare Pe Xa = 050} we increase and decrease this by an order of magnitude. Decreasing to $Pe = 0.5$ causes ethanol to rapidly diffuse out of the droplet, being depleted by $t = 2$, see figure \ref{compare Pe Xa = 050}(d). Contact line spreading is abruptly halted as solutal Marangoni stresses cease and the droplet begins to retract. With limited spreading, the droplet remains relatively thick with a spherical cap profile. Only water is present after $t = 2$ and so evaporation is predictably slow compared to superspreading cases. Increasing $Pe$ to \num{50} means ethanol is retained in the droplet for longer times. In this case it has the effect of maintaining the surface tension gradient from apex to contact line as well as the volatility of the mixture. We can see from figure \ref{compare Pe Xa = 050}(d) that ethanol is present in large concentrations at the apex until dry-out, suggesting it is also present in large concentration throughout the rest of the droplet. It is the retention of ethanol that results in higher evaporation rates over the interface and ultimately leads to faster evaporation and a shorter lifetime than the base case of $Pe = 5$. 
\subsection{Reynolds number}
\begin{figure}
	\centering
	\includegraphics[width=0.75\textwidth]{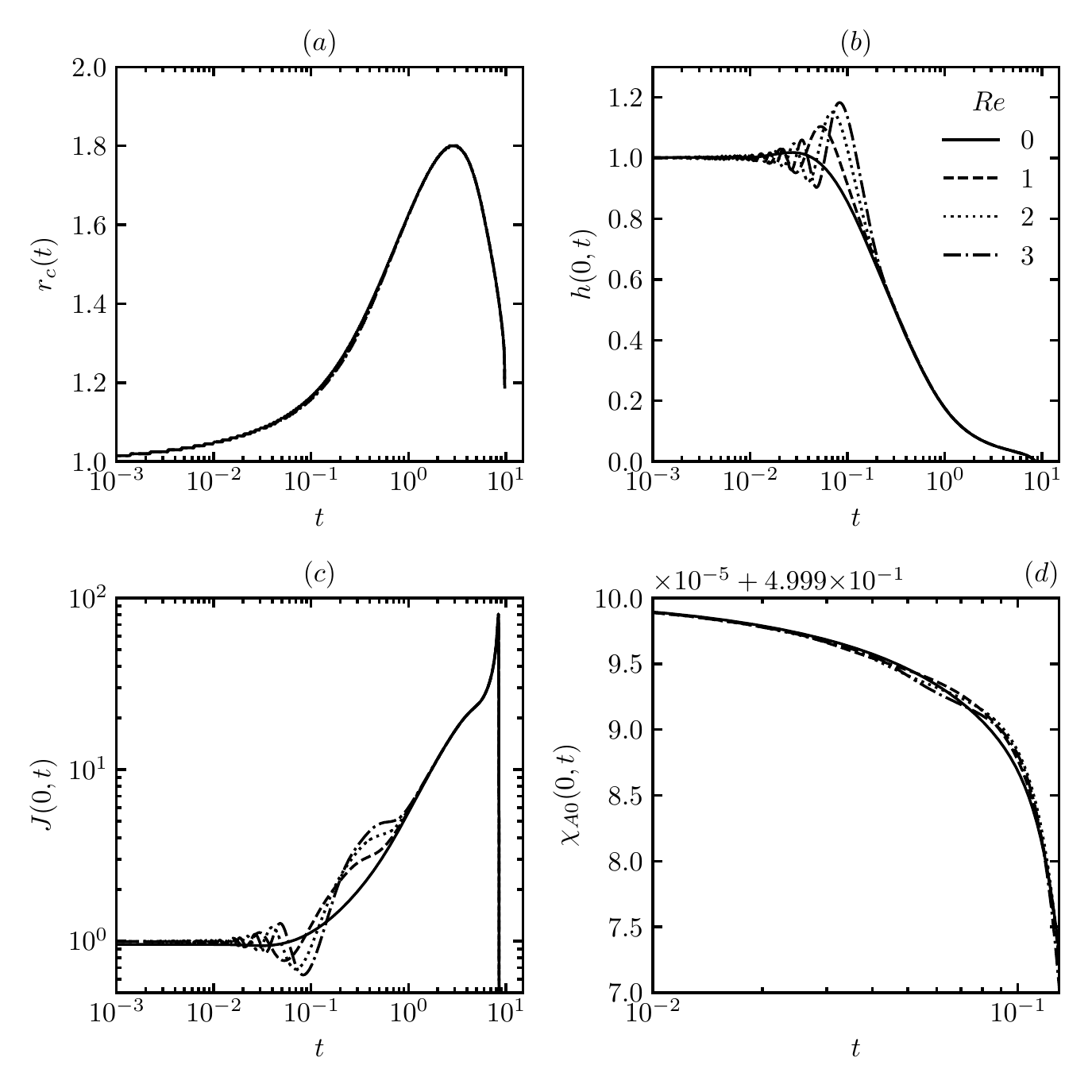}
	\caption{Profiles of (a) contact line position, (b) apex height, (c) apex mass flux, and (d) apex mass fraction throughout the lifetime of a $\chi_{A0,i} = 0.50$ droplet with varying Reynolds numbers, $Re$. Unless otherwise stated, dimensionless parameters are those given in table \ref{ethanol-water dimensionless base parameters table}.}
	\label{compare Re Xa = 050}
\end{figure}
Finally, we consider the effect of hydrodynamic flow by introducing inertia via the Reynolds number. As we have already shown in figure \ref{equal prop r h fig}, a non-zero $Re$ introduces oscillations in the interface profile near the apex at early times. The effect is found to be more dramatic in the binary ethanol-water droplet. In figure \ref{compare Re Xa = 050}, the Reynolds number is increased from $Re = 0$ to $Re = 3$. Figure \ref{compare Re Xa = 050}(a) indicates that this has little effect on the position of the contact line, however, the stronger hydrodynamic flow increases both the amplitude and frequency of the apex interface oscillations seen in figure \ref{compare Re Xa = 050}(b). Closer inspection of the evaporative flux and mass fraction in figure \ref{compare Re Xa = 050}(c) and (d) respectively reveal similar oscillations in these fields, also increasing in amplitude and frequency with $Re$.
\subsection{Comparison with experiments} \label{sec:compare_exp}
Given the nature of our one-sided model defined in section \ref{model definition}, we do not attempt a direct comparison to our experimental results presented in section \ref{experimental findings}. The lifetimes of experimental droplets are several orders of magnitude longer than our one-sided model predicts once a re-dimensionalisation is performed, although we could mitigate this somewhat by controlling $E$ and $K$, as shown in sections \ref{Evaporation number} and \ref{Knudsen number}. Evaporation could also be suppressed in our model by selecting a smaller accommodation coefficient in the Hertz-Knudsen expression, although this is not considered in the present study. The discrepancy between droplet lifetimes is not unexpected considering we use an accommodation coefficient of unity in our model while the experiments are performed under atmospheric air where, even at high substrate temperatures, diffusion of the vapour will play some role in evaporation. There are also additional effects of evaporative cooling and poor conductivity from the glass substrate in our experiments not accounted for in the model. Regardless, in their respective time frames, similar spreading rates (the same order of magnitude or closer) are predicted between the model and experiments, indicating that our one-sided model is sufficient to capture the main flow phenomena. The formation of a contact line ridge by our model at $\chi_{A0} = 0.50$ is very likely indicative of the beginning of the ``octopi'' patterns observed in the experiments as the same initial ethanol concentration. An obvious extension of this work would be to examine the effects of introducing significantly smaller accommodation coefficients to the evaporation model, likely providing a more favourable comparison to our experiments.
\section{Conclusions}
In surface tension dominated flows, whether they be planar layers of sessile droplets, the addition of a second miscible component introduces solutal Marangoni stress which can compete with or enhance the already present thermal Marangoni stress. With liquids comprising binary mixtures being a promising candidate for many modern micro cooling systems, it is essential these influences are understood. We have developed a one-sided model under the lubrication approximation to study the spreading and subsequent evaporation of volatile binary droplets consisting of an ethanol-water type mixtures deposited on a heated substrate. We considered specifically flat (low contact angle) droplets, assumed to be very thin such that their radius is much larger than their height. Droplets are released into precursor film, resulting in a freely moving effective contact line. Additionally, we conducted an experimental investigation into ethanol-water droplets deposited on heated borosilicate glass substrates with a hydrophilic coating to encourage spreading, similar to the conditions in our numerical model. An apparatus was designed to capture the droplets from above in an aerial viewpoint and a detection algorithm written to measure position of the contact line during spreading and recession. 

Experimentally, we investigated \SI{1}{\micro\litre} volumes of ethanol-water droplets comprising \SI{11}{\wtpercent}, \SI{25}{\wtpercent}, and \SI{50}{\wtpercent} initial ethanol concentration. The effect of increasing substrate temperature for \SI{30}{\celsius} to \SI{50}{\celsius} to \SI{70}{\celsius} on droplets comprising \SI{50}{\wtpercent} initial ethanol was also considered. We found that in all cases increasing initial ethanol concentration, and hence the magnitude of solutal Marangoni stresses, enhanced droplet spreading. This led to faster spreading rates while reducing the length of the spreading phase, resulting in a slightly reduced maximum droplet radius and shorter overall droplet lifetime. When initial ethanol concentration reached \SI{50}{\wtpercent}, a contact line instability emerges in the form of advancing fingers in an ``octopi'' arrangement accompanied by a second instability showing spoke-like patters arranged radially over the interface. Instabilities persist at all substrate temperatures for initial ethanol concentration of \SI{50}{\wtpercent}. The enhanced spreading rates cause the droplet interior to dry out before the contact line, leaving a ring where the contact line instability was previously present. The measured spreading rates closely match those predicted by our one-sided model in their respective time frames. The formation of the contact line ridge we observed in \SI{50}{\wtpercent} initial ethanol droplets preceding instability is also predicted by our model at the same concentration.

From a theoretical point of view, we have developed a numerical model and examined in detail the effect of increasing initial ethanol mass fraction in a binary ethanol-water droplet. We demonstrated the delicate interplay between solutal effects driving the droplet outwards and the competing thermal Marangoni stress encouraging the contact line to contract inward. With increasing strength of solutal Marangoni stress spreading rates, in some cases, were found to be compatible to those of superspreading surfactants such as trisiloxanes. In these cases, a ridge in the interface profile is formed ahead of the contact line, causing a thicker rim of liquid at the droplet edge rich in the less volatile component. This results in the droplet interior drying out before the edge, leaving the ridge to remain in the final stages of evaporation. This behaviour is similar to that seen in the alkane mixtures studied by \citet{Guena2007b}. We observed the same qualitative behaviour by our experiments. We then went on to conduct a parametric study, investigating the effects of other important parameters significantly affecting droplet behaviour. These included the evaporation rate (via $E$ and $K$), thermal Marangoni stress (via $Ma$), solutal Marangoni stress (via $\sigma_R$), mass diffusion (via $Pe$), and inertial effects (via $Re$). Although we do not attempt a direct experimental comparison due to the one-sided nature of our model, similar spreading rates are shared between the model and experimental result, suggesting that our one-sided model is sufficient to capture the main flow phenomena. 
\section{Acknowledgements}
The authors gratefully acknowledge the supports received from  ThermaSMART project of European Commission (Grant no. EC-H2020-RISE-ThermaSMART-778104). GK acknowledges the support received by the SPREAD project of Hellenic Foundation for Research and Innovation and General Secretariat for Research and Technology (Grant no. 792).  
\section{Declaration of interests}
The authors report no conflict of interest.
\bibliography{Bibliography}
\bibliographystyle{jfm}
\end{document}